\documentclass[12pt,twocolumn]{article}

\usepackage{amsmath}
\usepackage{amssymb}
\usepackage[latin1]{inputenc}
\usepackage{amsfonts}
\usepackage[dvips]{graphicx}
\usepackage{fancyheadings}

\title{On the width and shape of gaps in protoplanetary disks\\ \vskip 2truecm}

\author{{\bf A. Crida$^1$, A. Morbidelli$^1$, F. Masset$^2$}\\
\ \\ \small
$^1$\ O.C.A., B.P. 4229, 06304 Nice Cedex 4, France\\
\small
$^2$\ UMR AIM, DSM/DAPNIA/SAp, Orme des Merisiers, CE-Saclay, 91191 Gif/Yvette Cedex, France\,;\\
\small
IA-UNAM, Apartado Postal 70-264, Ciudad Universitaria, Mexico City 04510, Mexico\\
\small Contact\,: \texttt{crida@obs-nice.fr}\\
ICARUS, in press (accepted on October, 11, 2005).
}
\date{}

\begin{document}


\maketitle

\section*{Abstract}
{\small
Although it is well known that a massive planet opens a gap in a
proto-planetary gaseous disk, there is no analytic description of the
surface density profile in and near the gap. The simplest approach,
which is based upon the balance between the torques due to the
viscosity and the gravity of the planet and assumes local damping,
leads to gaps with overestimated width, especially at low
viscosity. Here, we take into account the fraction of the gravity
torque that is evacuated by pressure supported waves. With a novel
approach, which consists of following the fluid elements along their
trajectories, we show that the flux of angular momentum carried by the
waves corresponds to a pressure torque. The equilibrium profile {of
the disk} is then set by the balance between gravity, viscous and
pressure torques. We check that this balance is satisfied in numerical
simulations, with a planet on a fixed circular orbit. We then use a
reference numerical simulation to get an ansatz for the pressure
torque, that yields gap profiles for any value of the disk viscosity,
pressure scale height and planet to primary mass ratio. {Those} are in
good agreement with profiles obtained in numerical simulations {
over a wide range of parameters}. Finally, we provide a gap opening
criterion that simultaneously involves the planet mass, the disk
viscosity and the aspect ratio.
}


\section{Introduction}

The dynamical evolution of planets in proto-planetary disks has become
a topic of renewed interest in the last decade, boosted by the
discovery of extra-solar planets, and in particular of hot
Jupiters. In fact, the observation of giant planets close to their parent
stars argues for the existence of effective mechanisms of planetary
migration, which can be found in the study of planet-disk
interactions. 

Several types of migration have been identified, depending on how the
planet modifies the local density of the disk. {Type I migration
occurs when the planet is not massive enough to significantly alter
the local density of the disk\,; the planet migrates inward with a
speed proportional to its mass (Ward, 1997)}. Type II migration
corresponds to the case where the planet is so massive that it opens a
clear gap in the disk\,; the migration then depends on the viscous
evolution of the disk (Lin and Papaloizou, 1986\,a,\,b). Type III (or
runaway) migration corresponds to planets with intermediate mass,
which do not open a {clear} gap, but only form a dip around their
orbits in the gas surface density profile\,; under some conditions,
their migration drift rate can grow exponentially, in a runaway
process (Masset and Papaloizou, 2003).

The modification of the {disk density} is the result of the
competition of torques exerted on the disk by the planet and {by} the
disk itself. More precisely, the planet gives some angular momentum to
the outer part of the disk, {while it} takes some from the inner part
(Lin and Papaloizou, 1979\,; Goldreich and Tremaine, 1980). In doing
so, it pushes the outer part of the disk outward and the inner part
inward, and {therefore} tends to open a gap. However, the {
internal evolution of the disk, which tends to spread the gas into the
void regions, opposes to} the opening of the gap.

\begin{figure}[t!]
\includegraphics[width=0.63\linewidth,angle=270]{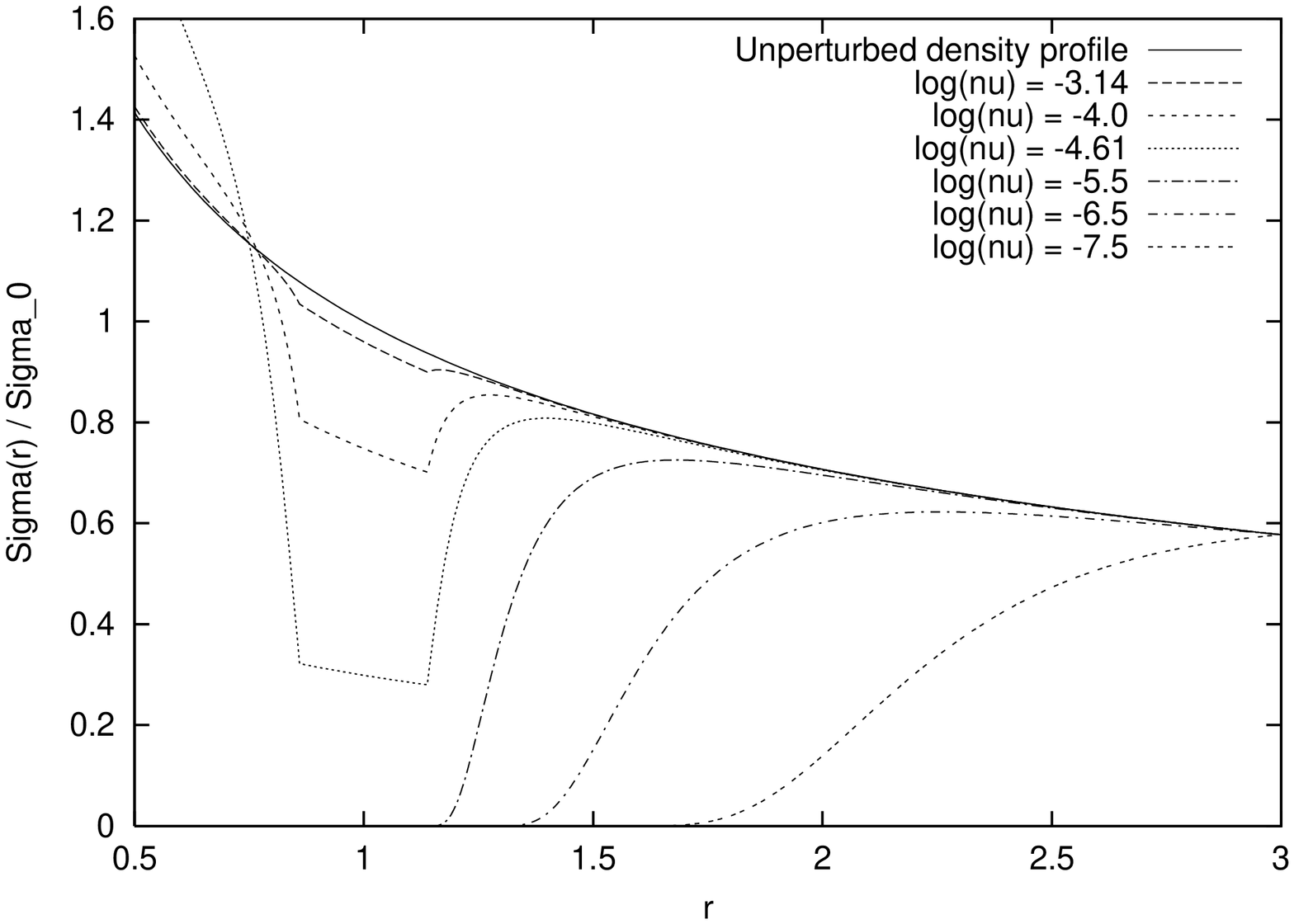}
\includegraphics[width=0.63\linewidth,angle=270]{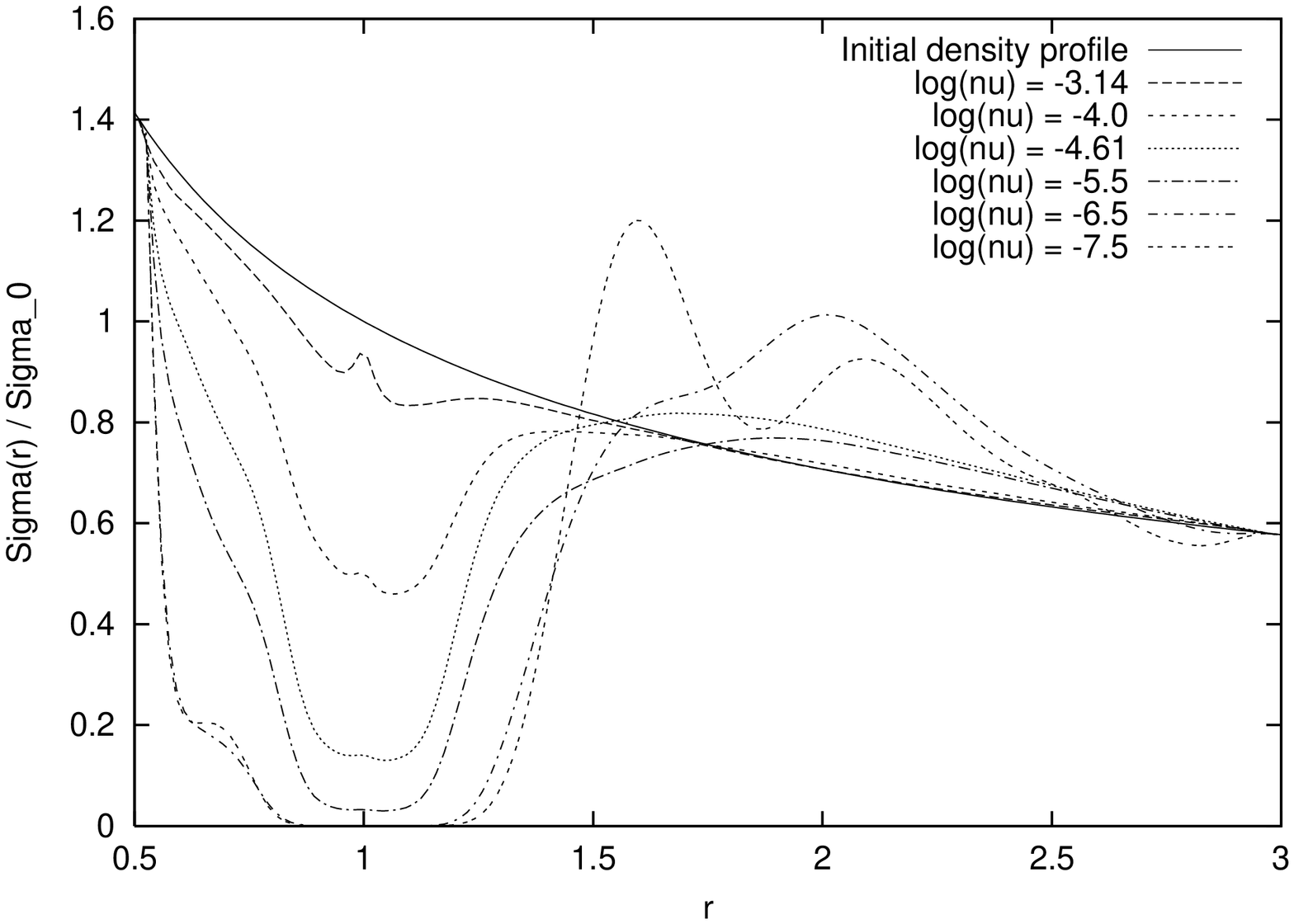}
\caption{\small Gap profiles created by a Jupiter mass planet, for
different viscosities. The vertical axis {represents} the normalized
azimuth-averaged density. The horizontal axis represents the distance
to { the} primary in normalized units. Top panel\,: analytic curves
obtained by matching the differential torques due to gravity and
viscosity on each { elementary ring of the disk}. Bottom panel\,:
numerical profiles obtained in simulations, after $1000$ planetary
orbits for the three largest viscosities, and $5000$ orbits for
$\log(\nu)=-5.5$ and $-6.5$. }
\label{fig:figure1poster}
\end{figure}

However, there is a lack of an analytical prediction of the gap
profiles. Classically, the gap is considered to have a step function
profile, with the edges located at the sites where the total gravity
torque is equal to the total viscous torque. This is obviously an
oversimplification. A more sophisticated approach has been recently
presented by Varni\`ere {\it et al.} (2004). They {provide an analytic
expression that} describes the gap profile, by equating the viscous
and gravity torques on any elementary disk ring. We will provide more
details on this approach in
section~\ref{sec:Gravity_and_Viscous_Torques}. The problem is that,
when the viscosity { is small}, the viscous torque { is small}
as well, and thus it cannot counterbalance the gravity
torque. Consequently, { in a low viscosity disk,} a non-migrating
planet should open{ a very wide gap, unlike what is observed} in
numerical simulations (see e.g. Fig.~\ref{fig:figure1poster})\,: gaps
{ do} increase in width and depth as the viscosity decreases, but
{ the dependence of the gap profile is less sensitive on viscosity
in the numerical simulations than it is expected in theory.}

The reason for this difference is that not all of the gravity torque
is locally deposited in the disk. It is transported away by density
waves (Goldreich and Nicholson, 1989\,; Papaloizou and Lin, 1984\,;
Rafikov, 2002{\,; see Appendix~C}). These waves are observed in
simulations. In this situation, the viscous torque has to
counterbalance only a {\it fraction} of the gravity torque, which
yields narrower gaps than expected from the simple
viscous/gravitational torque balance.

An evaluation of the fraction of the gravitational torque that is
locally deposited {at shock sites} in the disk has been undertaken by
Rafikov (2002). {However he did not use his analysis to provide an
analytic representation of the gap profile}. Moreover, his
calculation required several assumptions (the planet Hill radius had
to be much smaller than the local disk height, the disk surface
density was assumed to be uniform, etc.), which are not satisfied in
{ the general giant planet case}.

Here we introduce a novel approach. We follow a fluid element along
its trajectory which, in the steady state, is periodic in the planet
corotating frame. A flux of angular momentum carried by the density
waves corresponds to a pressure torque acting on the fluid element,
whose average over a synodic period is non-zero. In this work, we
evaluate this averaged pressure torque, together with the gravity and
viscous torques. The fact that the fluid element path is closed
implies that these time averaged torques { balance}.

In section~\ref{sec:Pressure_torque} we introduce the pressure torque,
and we { use numerical simulations to check} that the gap structure is
set by the equilibrium between the gravity torque and the sum of the
viscous and the pressure torques. In
section~\ref{sec:Semi-analytic_results} we construct a semi-analytic
algorithm and we get an expression to compute that gap profile. We
compare our results with the profiles obtained in numerical
simulation, and we discuss the merits and limitations of our
method. In section~\ref{sec:gap_profil_visc_AR} we use our algorithm
to explore the dependence of the gap structure on disk viscosity and
aspect ratio. We recover the trends observed in numerical simulations,
namely the { limited gap width in low viscosity disks} and the filling
of the gap with increasing viscosity and/or aspect ratio. Finally in
section~\ref{sec:new_criterion}, we provide a gap opening criterion
that simultaneously involves the viscosity, the scale height and the
planet mass.


\section{Gravity and Viscous Torques}

\label{sec:Gravity_and_Viscous_Torques}

{ In this section we revisit the calculation of the gravity and
viscous torques mentioned in the Introduction. We show that
considering them alone, as usually done, is not sufficient to achieve
a quantitatively correct description of the gap profiles.}

\subsection{Notations}

The disk is represented in cylindrical coordinates $(r,\theta,z)$,
centered on the star, where the plane $\{z=0\}$ corresponds to the
mid-plane of the disk. The disk viscosity $\nu$ and aspect ratio
$(H/r)$ --\,where $H$ denotes the thickness of the disk\,-- are
assumed to be invariant in time and space. The equations of fluid
dynamics are integrated with respect to the $z$-coordinate, so that
$z$ disappears from the equations and only two dimensions are
effectively used. This procedure introduces the concept of surface
density $\Sigma$, which is defined as $\int_{-H}^{+H}\rho\,{\rm d}z$,
where $\rho$ is the volume density in the disk.

In the theoretical analysis (but not in the numerical calculations)
the disk is { assumed} to be axisymmetric, so that $\Sigma$ only
depends on $r$. The angular velocity $\Omega$ is { assumed} to be
Keplerian\,: $\Omega\propto r^{-3/2}$. The planet is assumed on a
circular orbit around the star. The radius of its orbit is denoted
$r_p$. The mass of the planet is denoted $M_p$ and its ratio with the
mass of the central star $M_*$ is $q$. Normalized units are
introduced, so that $M_*=r_p=1$ and the gravitational constant $G$ is
also assumed to be unity. In the limit $q\to 0$, this sets the angular
orbital velocity of the planet $\Omega_p=1$ and its period equal to
$2\pi$.

\subsection{Total torques}
\label{Tt}

Usually, one considers the part of the disk extending from a given
radius $r_0>r_p$ to infinity. The study of the part of the disk
extending from 0 to $r_0<r_p$ is done in an analogous way. Two torques
are evaluated. The first one is due to the disk viscosity and can be
easily derived from the stress tensor in a Keplerian disk { with
circular orbits}. The torque exerted on the considered part of the
disk ($r>r_0$) by the complementary part is written (see for instance
Lin and Papaloizou, 1993)\,:
\begin{equation}
 T_\nu = 3 \pi \Sigma \nu {r_0}^2 \Omega_0
\label{eq:Tnu}
\end{equation}
{ Notice that more refined expression for perturbed disks with
eccentric orbits have been proposed in the literature (see for
instance Borderies {\it et al.} 1982), but they have not been used
in the works that we review in this section.}

The second torque comes from the gravity of the planet. { It can be
computed following two different approaches. In the first one
(Goldreich and Tremaine, 1980\,; Ward, 1986), it is decomposed into
the sum of the individual torques exerted at each Lindblad resonance.
In the second approach (Lin and Papaloizou, 1979\,; Goldreich and
Tremaine, 1980) it is obtained by computing the angular momentum
change for fluid elements at conjunction with the planet, using an
impulse approximation. The two approaches are known to give equivalent
results. In the following, we use the expression from the impulse
approximation\,:}
\begin{equation}
 T_g(\Delta_0) = C\,q^2 \Sigma {r_p}^4 {\Omega_p}^2
 \left(\frac{r_p}{\Delta_0}\right)^3\ ,
\label{eq:Tg}
\end{equation}
where $\Delta_0=(r_0-r_p)$. The above expression gives the torque
 exerted by the planet on a disk extended from $r_0$ to infinity. It
 is valid only for $|\Delta_0|>\Delta_{\rm m}$, where $\Delta_{\rm m}$
 is the maximum { of} $H$ (the local thickness of the disk) and the
 Hill radius of the planet $R_H=(q/3)^{1/3}$ (Goldreich and Tremaine,
 1980\,; Ward, 1997). The value of the numerical coefficient $C$
 depends on the approach followed for the calculation of the
 torque. In the most recent and refined calculation, Lin and
 Papaloizou (1993) found $C =
 \frac{32}{243}[2K_0(\frac{2}{3})+K_1(\frac{2}{3})]^2 \approx 0.836$
 (where $K_0$ and $K_1$ are modified Bessel functions).

Classically, the gap is modeled as a step function profile in the
disk surface density, with edges placed at a distance $\Delta_0$ from
the planet orbit, with $\Delta_0$ given by the solution of the
equation $T_g(\Delta_0)=T_\nu$, and $T_g$ and $T_\nu$ given in
\eqref{eq:Tg} and \eqref{eq:Tnu}. The maximal gravity torque is\,: 
\begin{equation*}
 T_g(\Delta_{\rm m}) \approx 0.836\,q^2 \Sigma {r_p}^4
 {\Omega_p}^2\left(\frac{r_p}{\Delta_{\rm m}}\right)^3\ .
\end{equation*}
Thus, a gap can be opened only if\,:
\begin{equation}
\nu < 0.0887 q^2 \frac{{r_p}^4{\Omega_p}^2}{{r_0}^2\Omega_0}
\left(\frac{r_p}{\Delta_{\rm m}}\right)^3\ ,
\label{eq:nucrit}
\end{equation}
otherwise $T_\nu$ is larger than $T_g$ and the gas overruns the
planet. Condition \eqref{eq:nucrit} is equivalent to that given in
Bryden \textit{et al.} (1999), expressed as a constraint on the mass
of the planet relative to the viscosity of the disk.

\subsection{Differential torques and comparison with numerical tests}

\paragraph{Differential torques.}
Varni\`ere {\it et al.} (2004) proposed a more refined approach to
model the surface density profile in the gap. Their approach is based
on a simple consideration\,: in equilibrium, when a steady state
is reached, the gravity torque and the viscous torque must be equal on
every elementary ring of the disk. The torques acting on
elementary rings can be computed by differentiation relative to
$r\equiv r_0$ of \eqref{eq:Tnu} and \eqref{eq:Tg}\,:
\begin{equation}
 \delta T_\nu(r) = -\frac{3}{2} \nu \Omega
 \left[\frac{r}{\Sigma}\frac{{\rm d}\Sigma}{{\rm d}r}+\frac{1}{2}\right]\left(2\pi
 r \Sigma\right)
\label{eq:dTnu}
\end{equation}
\begin{equation}
\delta T_g(r) \approx 0.4\,q^2 {r_p}^3 {\Omega_p}^2 r^{-1}
\left(\frac{r_p}{\Delta}\right)^4 \left(2\pi r \Sigma\right)
\label{eq:dTg}
\end{equation}
Matching $\delta T_\nu$ and $\delta T_g$ gives a differential equation
in $\Sigma$\,:
\begin{equation}
\frac{1}{\Sigma}\frac{{\rm d}\Sigma}{{\rm d}r}=\frac{\delta T_g(r)}{3\pi\nu {r}^2 \Omega \Sigma}-\frac{1}{2r}
\label{eq:eqdiff}
\end{equation}
The integration of this equation gives the profile of the gap. 

The top panel of Fig.~\ref{fig:figure1poster} gives examples { of
the solution of Eq.~\eqref{eq:eqdiff} }for several values of the
viscosity, from strong ($\nu=10^{-3.14}$) to weak
($\nu=10^{-6.5}$). To compute them from \eqref{eq:eqdiff}, we have (i)
assumed that the mass of the planet is $10^{-3}$ in our normalized
units, (ii) imposed the boundary condition $\Sigma(r_0=3)=1/\sqrt{3}$
and (iii) assumed that the gravity torque is null in { the
horseshoe region, here approximated by\,:}
$r_p-2\,R_H<r<r_p+2\,R_H$. As a consequence of (iii) the surface
density profile in the vicinity of the planet assumes an equilibrium
slope proportional to $1/\sqrt{r}$, which makes $\delta T_\nu$
null. Notice that the slopes of the surface density at the edges of
the gap do not depend on our assumptions (ii) and (iii), but are
dictated solely by the differential equation \eqref{eq:eqdiff}. {
We remark that the profiles illustrated in the figure are the same as
in Varni\`ere {\it et al.} 2004, despite the fact that these authors
consider the gravity torque as given by the sum of the individual
Lindblad resonances. This again underlines the equivalence of the two
approaches for the calculation of the gravity torque discussed in
\ref{Tt}.}

\paragraph{Numerical simulations.}
We have tested the results of these analytic calculations using purely
numerical simulations. For this purpose, we have used the 2D
hydrodynamic code described in Masset (2000), and considered a Jupiter
mass planet ($q=10^{-3}$) in a disk, whose initial surface density
profile decays as $1/\sqrt{r}${, and $\Sigma(r_p)=6.10^{-4}$ (the
value of the minimal mass solar nebula at 5 AU (Hayashi, 1981))}. The
disk aspect ratio was fixed at $5\%$. The viscosity was chosen equal
to the values used for the analytic computations, for direct
comparison. In these simulations, the planet was assumed not to feel
the gravity of the disk, so that it did not migrate. The grid used by
the code for the hydrodynamical calculations extended from $0.5$ to
$3$ (we remind that the planet location is $r_p=1$). { The boundary
conditions in $r$ are non-reflecting, which means that the waves
behave as if they were propagating outside the boundaries of the
grid. The angular momentum they carry is thus lost\,; we have checked
that the flux through the outer boundary represents only a negligible
fraction of the total gravity torque (see Appendix~C). The relative
surface density amplitude perturbation at the outer boundary has been
measured to be less than 5\%.} The size of the grid was 150 cells in
radius and 325 cells in azimuth. The simulations were carried on for
$1000$ planetary orbits, for viscosity down to $10^{-5.0}$ and $5000$
orbits for weaker viscosities. At these times, the profile of the gap
{ does not seem to} evolve significantly any more, as also found by
Varni\`ere {\it et al.} (2004).

\paragraph{Comparisons.}
The results are illustrated in the bottom panel of
Fig.~\ref{fig:figure1poster}. As anticipated in the introduction, we
remark an evident difference with the analytic predictions. The
simulated gap is much narrower than the one predicted by the analytic
expression \eqref{eq:eqdiff} for low viscosities ($\nu<10^{-6}$).
At first sight, one might think that the discrepancy between { the
analytical and numerical solutions} is due to the numerical viscosity
{(dissipation due to numerical errors)} of the computer
code. However, this is unlikely for the following reasons\,:
(i) different gap profiles are observed for different viscosities,
which shows that the simulation is not dominated by the numerical
viscosity, as the latter should be the same in all simulations\,; (ii)
changing the resolution of the grid used in the numerical scheme,
which changes the numerical viscosity, does not affect the gap
profiles significantly\,; (iii) different numerical schemes give
consistent results (De Valborro, private communication).

As anticipated in the introduction, the problem with this {
analytical} modeling is the assumption that the gravity torque is
entirely deposited in each annulus of the disk. A condition for {
such} deposition to happen is that $R_H\gtrsim H$ (Lin and Papaloizou,
1993). Thus, this is usually considered as a second independent
criterion for gap opening, in addition to \eqref{eq:nucrit} (Bate {\it
et al.}, 2003).  However, even if this condition is satisfied, a
fraction of the gravity torque is still evacuated by the waves
(Goldreich and Nicholson, 1989\,; Papaloizou and Lin, 1984\,; Rafikov,
2002). The problem is to evaluate this fraction. Below we show that it
can be computed from a mean pressure torque acting on the fluid
elements over their periodic equilibrium trajectories.


\section{Pressure torque}

\label{sec:Pressure_torque}

\begin{figure}[t!]
\includegraphics[width=0.7\linewidth,angle=270]{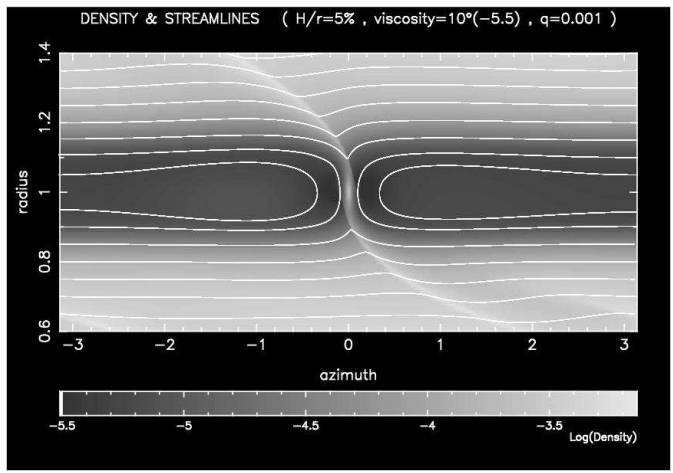}
\caption{\small Disk surface density map in the vicinity of a gap
opened by a Jupiter mass planet located at $(r_p=1,\theta_p=0)$. Light
grey denotes high density and black low density, in a logarithmic
scale. The white curves show some {streamlines}, in the frame
corotating with the planet. They are followed from $\pi$ to $-\pi$ for
$r>1$, and from $-\pi$ to $\pi$ for $r<1$, periodically. Two of
them correspond to horseshoe orbits in the planet corotation
region. Notice the strong distortion of the streamlines when
they cross the over-density corresponding to the spiral wave
(wake) launched by the planet.}
\label{fig:wake}
\end{figure}

{Consider an arbitrary closed curve in the disk and a little tube
around it. The rate of change of angular momentum of the matter in
the tube is the sum of the differential flux of angular
momentum through its boundaries (due to the advection of matter) and
of the torques acting on it. From the Navier-Stokes equations, in
addition to the gravity and viscous torques, there is a third torque
due to pressure\,:
\begin{equation}
t_P= \oint {{c_s}^2}\frac{\partial\Sigma}{\partial\theta}{\rm d}\theta\ ,
\label{eq:tPtube}
\end{equation}
where the integral is computed along the curve and we have assumed the
usual state equation $P={c_s}^2\Sigma$ (with $c_s=H\Omega$ denoting
the sound speed). If the tube is a ring centered at the origin $r=0$,
the torque $t_P$ is equal to zero (on the ring, $r$ is constant and
$\partial\Sigma/\partial\theta={\rm d}\Sigma/{\rm d}\theta$), while
the differential flux of angular momentum is generally not zero. The
latter is the flux carried by the pressure supported wave (Goldreich
and Nicholson, 1989{\,; see Appendix~C}). On the contrary, if one
chooses a stream tube ({\it i.e.} a tube bounded by two neighboring
streamlines), the differential flux is obviously zero (there is no
flux of matter, by definition of stream tube), while $t_P$ is non zero
in general. The latter is true because the streamlines are strongly
distorted (see Fig.~\ref{fig:wake}) so that $\Sigma\equiv
\Sigma(r(\theta),\theta)$ and thus $
\partial\Sigma/\partial\theta\neq{\rm d}\Sigma/{\rm d}\theta$. This
shows that one can translate the angular momentum flux carried by the
waves into a pressure torque, by a suitable partition of the disk in
concentric tubes. Obviously the two approaches are equivalent as the
physics is the same. However, working with stream tubes and pressure
torques gives practical computational advantages. This is therefore
the approach that we follow in this paper.

Below, we check that the gravity, pressure and viscous torques really
cancel each other in the numerical simulations, once the steady state
is reached.}

\subsection{Computation of the torques along the trajectories}
\label{sub:Computation_torques}

The approach outlined above requires that the stream tubes are
closed. In our simulations this is true at the steady state, because
our boundary conditions preserve the initial radial velocity at the
edges of the grid, which is null as a result of our choice for the
initial disk density profile $\Sigma \propto 1/\sqrt{r}$. The
calculation of the streamlines, which is done in Fourier space to
ensure periodicity, is detailed in Appendix~A. We remark that, in the
steady state, the streamlines coincide with the fluid element
trajectories.

Denoting the streamline by $r_i(\theta)$, we numerically compute the
following expressions, { which are the integrals of
$(1/\Sigma)(r\textbf{F}_\theta)$ with $\textbf{F}_\theta$ the
azimuthal component of the force due to gravity, viscosity, or
pressure respectively}\,:
\begin{equation}
\label{eq:good_torques_tg}
t_g(\theta)  =  \frac{1}{T_i(\theta)} \int_\pi^{\theta} r_i(\theta')\frac{\partial\phi_{(r_i(\theta'),\theta')}}{\partial\theta'} \left|\frac{d\theta'}{\dot{\theta'}}\right|
\end{equation}
\begin{equation}
\label{eq:good_torques_tnu}
t_\nu(\theta)  =  \frac{1}{T_i(\theta)} \int_\pi^{\theta} \frac{1}{\Sigma_{(r_i(\theta'),\theta')}}r_i(\theta')\textbf{F}_\theta^\nu(r_i(\theta'),\theta') \left|\frac{d\theta'}{\dot{\theta'}}\right|
\end{equation}
\begin{equation}
t_P(\theta)  =  \frac{1}{T_i(\theta)} \int_\pi^{\theta} \frac{{c_s}^2}{\Sigma_{(r_i(\theta'),\theta')}}\frac{\partial\Sigma_{(r_i(\theta'),\theta')}}{\partial\theta'}\left|\frac{d\theta'}{\dot{\theta'}}\right|
\label{eq:good_torques_tP}
\end{equation}
Here, $\phi$ denotes the gravitational potential of the planet, {
and $\textbf{F}_\theta^\nu=\frac{1}{r}\left[\frac{\partial}{\partial
r}\left(r\bar{\bar{T}}_{r\theta}\right)+\frac{\partial}{\partial
\theta'}\bar{\bar{T}}_{\theta\theta}+\bar{\bar{T}}_{r\theta} \right]$,
where $\bar{\bar{T}}=\left(
\begin{array}{cc}
\bar{\bar{T}}_{rr} & \bar{\bar{T}}_{r\theta}\\
\bar{\bar{T}}_{\theta r} & \bar{\bar{T}}_{\theta\theta}
\end{array}
\right)$ is the local viscous stress tensor for a Newtonian fluid\,:
$\bar{\bar{T}}=2\Sigma\nu \left(\bar{\bar{D}}-(\frac{1}{3}\nabla
\vec{v})I\right)$, where $\bar{\bar{D}}$ is the strain tensor and $I$
is the identity matrix.}  We integrate from $\pi$ to $\theta$, with
$\pi>\theta\geqslant-\pi$, because we consider $r_0>r_p$, so that the
angular velocity is negative in the corotating frame. The time
required to reach $\theta$ from $\pi$ is denoted $T_i(\theta)$. Thus,
as the trajectories coincide with the streamlines in the steady state,
the expressions above describe the averaged torques felt by a fluid
element that travels from the planet opposition to $\theta$.

In the following, we denote for simplicity by $t_g$, $t_\nu$, $t_P$
the expressions \eqref{eq:good_torques_tg}-\eqref{eq:good_torques_tP}
evaluated at $\theta=-\pi$. The total torques acting on the stream
tube centered around the considered streamline are simply the product
of these quantities { times} the mass carried by the tube.



\begin{figure}[t!]
\includegraphics[width=0.7\linewidth,angle=270]{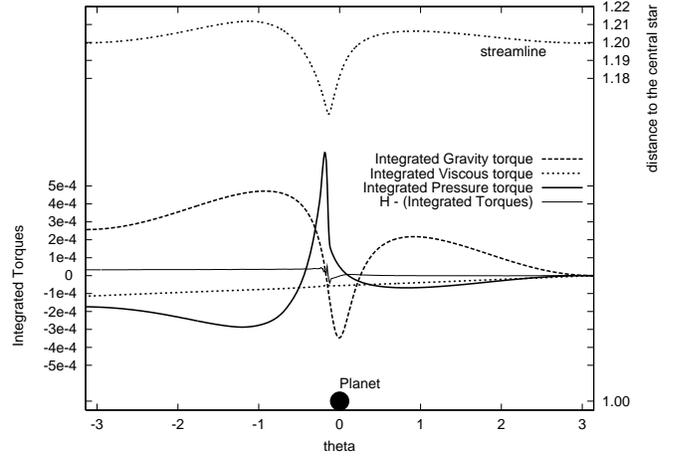}
\caption{\small Graphical representation of the expressions
\eqref{eq:good_torques_tg} (integrated gravity torque, bold short-dash
curve), \eqref{eq:good_torques_tnu} (integrated viscous torque, bold
dotted curve) and \eqref{eq:good_torques_tP} (integrated pressure
torque, bold solid curve). Their reference scale is reported on the
left vertical axis. The streamline followed for their calculation is
plotted in the planet corotating frame as a dashed curve at the top of
the figure { and can also be seen on Fig.~\ref{fig:wake}}, while the
position of the planet is shown by a filled dot at the bottom\,; the
corresponding scale is reported on the right vertical axis. The thin
solid curve shows the difference between the angular momentum measured
along the streamline ($H(\theta)$ ), and the sum of the three
integrated torques and of the initial angular momentum (${\cal
H}(\theta)$ ). A small difference is almost impulsively acquired at
the wake crossing, due to numerical approximations.}
\label{fig:torquesC}
\end{figure}

In the {next paragraphs} we give a brief description of the integrated
torques \eqref{eq:good_torques_tg}-\eqref{eq:good_torques_tP} as
functions of $\theta$, which are plotted { in
Fig.~\ref{fig:torquesC} for the streamline starting at $r=1.2$ at
opposition with the planet.}

\paragraph{viscous torque\,:}
The growth of the integrated viscous torque appears to be nearly
linear with respect to the azimuth, leading to a total negative
torque. We verified that { on this streamline} $t_\nu\approx\delta
T_\nu/(2\pi\Sigma r)$, with $\delta T_\nu$ { from}
\eqref{eq:dTnu}. Thus, the viscous torque depends only on the radial
relative derivative of the azimuthally averaged density
$\frac{1}{\Sigma}\frac{{\rm d}\Sigma}{{\rm d}r}$. { However, on
streamlines that pass closer to the planet, the difference between
$t_\nu$ and $\delta T_\nu/(2\pi\Sigma r)$ becomes more significant
(see Fig.~\ref{fig:TpTgTnu_r0}).}

\paragraph{gravity torque\,:}
The evolution of this integrated torque is not monotonic. The fluid
element is first repelled by the planet, as a result of the indirect
term in the gravitational potential. Then, when $\theta$ decreases
below $\sim 0.5$, it starts to be attracted by the planet. The
attraction becomes stronger and stronger as the fluid element
approaches conjunction, namely as $\theta$ decreases to $0$. The
integrated torque becomes negative. After conjunction, the planet
tends to pull the fluid element toward positive $\theta$, giving a
positive torque. As a result, the fluid element is rapidly repelled
toward larger $r$, as one can see { from the trajectory on
Fig.~\ref{fig:torquesC}}. This is typical of the scattering of test
particles in the restricted three body problem, which qualitatively
justifies the impulse approach for the calculation of the
gravitational torque, as in Lin and Papaloizou (1979).

\begin{figure}[t!]
\includegraphics[width=0.7\linewidth,angle=270]{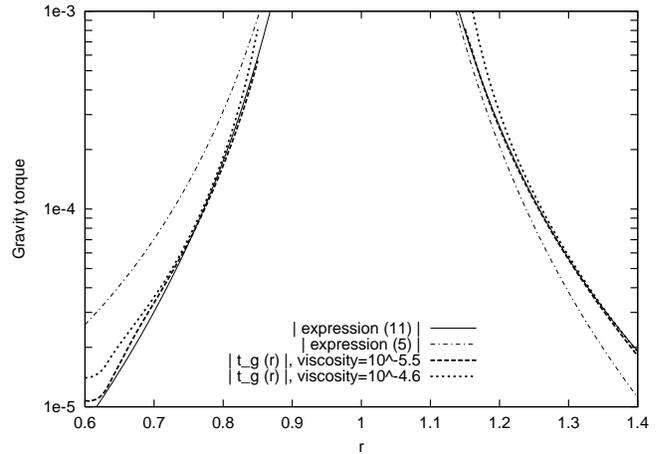}
\caption{\small Absolute value of $t_g$ as a function of $r$, where $r$
is the distance to the central star at planet opposition of the
streamline on which the gravity torque is integrated. The bold
long-dashed curve is obtained from the simulation with $q=10^{-3}$,
$(H/r)=5\%$, and $\nu=10^{-5.5}$, while the bold short-dashed curve
corresponds to a more viscous case ($\nu=10^{-4.6}$). The solid line
traces expression \eqref{eq:t_g}, which remarkably fits the results
obtained in the less viscous case. The dashed-dotted curve traces
expression \eqref{eq:dTg}, which shows that $\delta T_g\neq t_g$.}
\label{fig:gravity_torque}
\end{figure}

However, Lin and Papaloizou's calculation holds in the approximation
$r\!\sim\!r_p$. By comparing the numerical estimate of $\delta
T_g/(2\pi\Sigma r)$ with $\delta T_g$ given by Eq.~(\ref{eq:dTg}), we
find that the following expression, which has the same dependence in
$\Delta$ and nearly the same numerical coefficient, but which
distinguishes $r$ and $r_p$, provides a much more accurate
representation of the gravity torque\,:
\begin{equation}
t_g = 0.35\ q^2 {r_p}^{\!5}\,{\Omega_p}^{\!2}\,r \left(\frac{1}{\Delta}\right)^{\!4} {\rm sgn}(\Delta)\ ,
\label{eq:t_g}
\end{equation}
In reality, $t_g$ depends on the exact shape of the streamlines, which
in turn depends on the scale height and the viscosity (see
Fig.~\ref{fig:gravity_torque} and \ref{fig:TpTgTnu_r0}). However,
the difference is moderate and limited to the vicinity of the planet,
so that in the following we use expression \eqref{eq:t_g} for all
cases.

{ We stress that expression \eqref{eq:t_g} gives the torque exerted
on the fluid element, which is generally {\it not} the torque
deposited in the disk. In fact, even in the absence of viscosity, it
does not correspond to the change of angular momentum of the fluid
element, because some of the angular momentum is carried away by the
pressure torque.}

\begin{figure}[t!]
\includegraphics[width=1.35\linewidth]{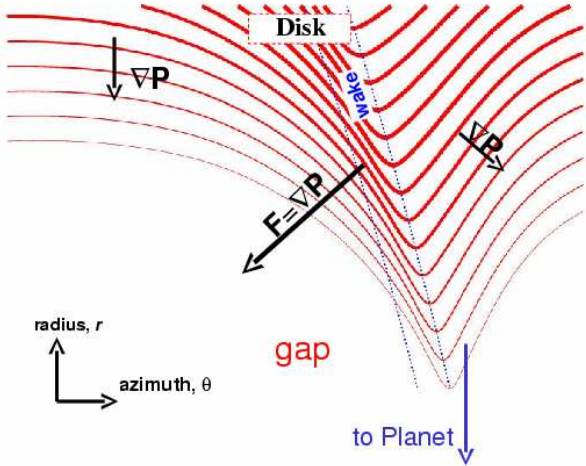}
\caption{\small Sketch on the origin of the pressure torque. Here are
drawn some streamlines, the width of which represents the mass {
carried by the corresponding streamtube}. At the gap edge, the
pressure gradient gives a force.  The distortion of the streamlines at
the wake leads to a large azimuthal component of this force, which
gives a torque.}
\label{fig:gradient}
\end{figure}

\paragraph{pressure torque\,:}
The dependence of this torque on $\theta$ is simple to understand if
one takes into account that\,: (i) the trajectories cross the wake
immediately after the conjunction with the planet {
(Fig.~\ref{fig:wake})}\,; (ii) the wake is a strong over-density in
the disk\,; (iii) the pressure term $\partial\Sigma/\partial\theta$
makes over-densities repellent. Thus, as the fluid element approaches
the wake, its azimuth $\theta$ decreasing in the corotating frame, the
pressure rises and tends to push the fluid element back in the
direction of increasing $\theta$.  This gives a positive local torque
and it explains the peak in the integrated pressure torque in
Fig.~\ref{fig:torquesC}. Then, after that the fluid element has
crossed the wake, the pressure decreases as $\theta$ decreases. This
leads to a negative local pressure torque. It corresponds to the fall
after the peak on Fig.~\ref{fig:torquesC}.  The negative contribution
is bigger than the positive one because of the asymmetry of the
trajectory relative to the wake position, which is clearly visible in
Fig.~\ref{fig:gradient}.

Clearly, the pressure torque must depend on the shape of the
streamlines and on the surface density relative radial gradient, which
govern the shape of the wake and its density enhancement. We return to
this in section~\ref{sec:Semi-analytic_results}.

\subsection{Torque balance at equilibrium}
\label{sub:Torque_balance}

From Fig.~\ref{fig:torquesC} we remark that, at $\theta=-\pi$,
the sum of the viscous and pressure torques is basically the opposite
of the gravity torque. Therefore, the three torques approximately
balance out.

{Given the angular momentum $H$ of a fluid element at $\theta=\pi$,
one can compute the angular momentum ${\cal H}(\theta)$ that it would
have if its trajectory were governed exclusively by the three torques
mentioned above\,: ${\cal H}(\theta)=
H(\pi)+t_g(\theta)+t_\nu(\theta)+t_P(\theta)$. This can be compared
with the local angular momentum on the trajectory $H(\theta)$,
measured directly from the numerical simulation.} In
Fig.~\ref{fig:torquesC} the thin line shows
$H(\theta)-\mathcal{H}(\theta)$. This function is zero for $\theta$
evolving from $\pi$ down to $\sim 0$, where the wake is crossed. At
the wake crossing, a small kick is observed. Then, when $\theta$
evolves from the wake location to $-\pi$, the function $H(\theta)-
\mathcal{H}(\theta)$ remains constant again. This confirms that
the trajectory is essentially governed by the three torques mentioned
above.

The small difference between $H$ and ${\cal H}$
(Fig.~\ref{fig:torquesC}) could in principle be due to the
pseudo-viscous pressure introduced in the simulation to avoid
numerical instabilities (Lin and Papaloizou, 1986a), but we have
verified that the effect of the latter is negligible. Thus, we
conclude that it is a consequence of numerical errors, introduced by
the grid discretization at the shock site.  This numerical issue
evidently prevents the three cumulative torques from balancing out
perfectly at $\theta=-\pi$\,: indeed, their sum is equal to
$\mathcal{H}(-\pi)-H(-\pi)\ne 0$.

\begin{figure}[t!]
\includegraphics[width=0.63\linewidth,angle=270]{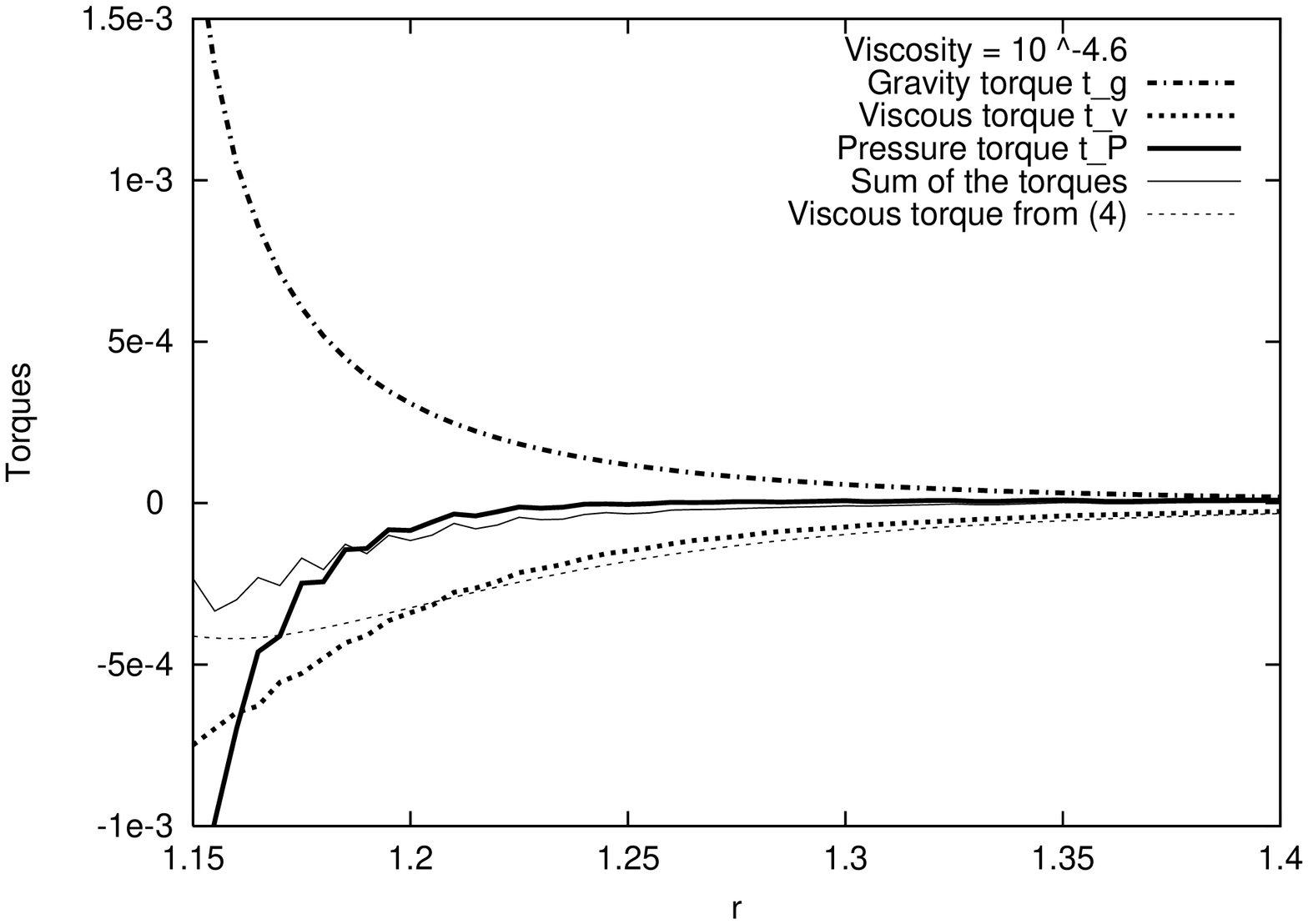}
\includegraphics[width=0.63\linewidth,angle=270]{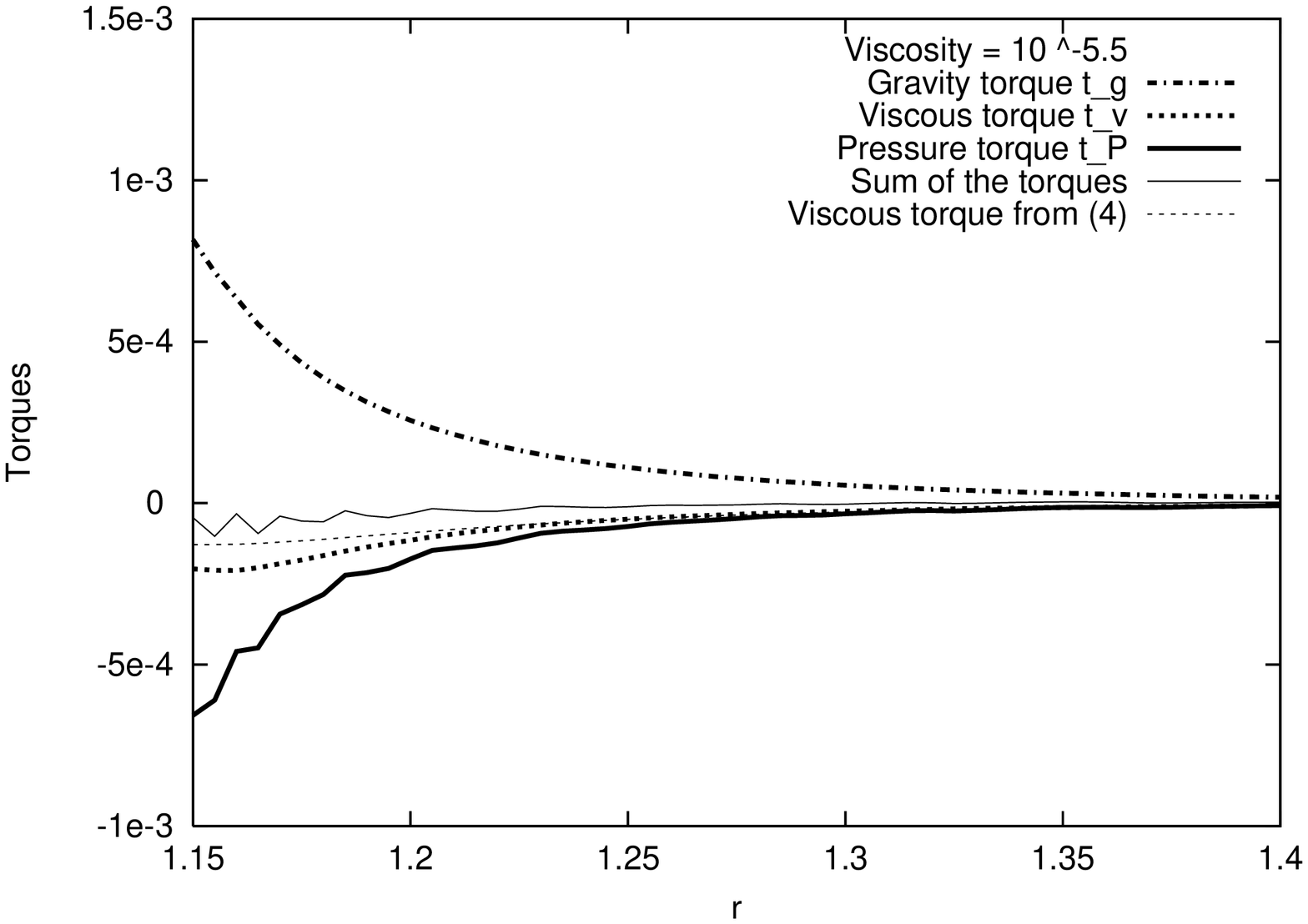}
\caption{\small The torques $t_g$, $t_\nu$, $t_P$ are plotted in bold lines as a
function of $r$, which denotes here and in the following plots the
radius of the streamline at opposition with respect to the planet
($\theta=\pm\pi$)\,; the horseshoe region $r\lesssim 1.15 \sim r_p+2\,R_H$ (see
Fig.~\ref{fig:wake}) is excluded. The thin dotted line shows the value
of the viscous torque given by $\delta T_\nu/(2\pi\Sigma r)$, with
$\delta T_\nu$ from \eqref{eq:dTnu} (keplerian circular
approximation). The thin solid line is the sum of the three torques.
It is not exactly zero, in particular in the vicinity of the planet,
because of numerical approximations generated at the wake
crossing. Top panel\,: large viscosity case\,; the pressure torque
becomes relevant only close to the planet. Bottom panel\,: low
viscosity case\,; the pressure torque appears further from planet,
compensating for the smaller viscous torque.}
\label{fig:TpTgTnu_r0}
\end{figure}

In order to explore the relative importance of viscosity and pressure
in different situations, we show in Fig.~\ref{fig:TpTgTnu_r0} the
three averaged torques as a function of $r$ for two simulations, with
$\nu=10^{-4.6}$ (top panel) and $\nu=10^{-5.5}$ (bottom panel). In the
more viscous case, the pressure torque becomes relevant for $r<1.2$,
{\it i.e.} at the edge of the gap. There, it substantially helps the
viscous torque in counterbalancing the gravity torque. This explains
why the gap observed in the simulation is narrower than the one
predicted by the theory considering only the gravity and the viscous
torques alone (see Fig.~\ref{fig:figure1poster}). In fact, if the
pressure torque were not present, all over the region $r<1.2$ the
relative radial gradient of the surface density of the disk would have
needed to be much steeper, in order to enhance the viscous torque up
to the value of the gravity torque (see Eq.~\eqref{eq:dTnu}). This would
have { given} a wider and deeper gap { profile}.

It is interesting to compare the top panel of
Fig.~\ref{fig:TpTgTnu_r0} with the lower panel, which is plotted for a
value of the viscosity that is almost an order of magnitude
smaller. First, we remark that the gravity torque is somewhat smaller
in the vicinity of the planet\,; this is due to a (moderate) change of
the shape of the streamlines, as discussed in last subsection.  The
viscous torque has decreased much more than the gravity torque, but
not proportionally to the viscosity\,; this is because the profile of
the gap has changed and the relative radial gradient of the surface
density is now steeper. The pressure torque has {\it increased}
relative to the gravity torque, and is now non-negligible in the full
region $r<1.3$. It is always larger in absolute value than the viscous
torque. Its radial profile looks very similar to that of the gravity
torque. In essence, it is the pressure torque { that} counterbalances
the action of the planet, with the viscosity only playing a minor
role. Thus there is a dramatic qualitative change, with respect to the
previous case, in how the torques balance out to settle the
equilibrium configuration.

The two cases discussed above convincingly show that the disk
equilibrium is set by the equation
\begin{equation}
t_g+t_\nu+t_P=0\ .
\label{eq:sum=0}
\end{equation}
When the viscosity fades, the role of pressure takes over in
controlling the gap opening process, limiting the gap width. This
means that, as viscosity decreases, a larger fraction of the gravity
torque is transported away by the pressure supported waves. This
phenomenon explains why the width of the gap increases with decreasing
viscosity in a much less pronounced way than in Varni\`ere {\it et
al.}'s model{, which does not include a pressure torque}.

\vskip 15pt

The role of pressure in limiting the gap width may still appear
surprising, but it can be understood with some physical
intuition. { In} an inertial environment, it is pressure
--\,and not viscosity\,-- which makes a gas fill the void space. In a
rotating disk the situation is different, because a radial pressure
gradient simply adds or subtracts a force to the gravitational force
exerted by the central star. This changes the angular velocity of
rotation of the gas, without causing any radial transport of
matter. Thus if the edges of the gap were circular, the pressure could
not play any role in limiting the gap opening. However, as the gap
edges are not circular, as shown in Fig.~\ref{fig:gradient}, the
pressure gradient is not entirely in the radial direction, and thus it
exerts a force with a non-null azimuthal component. This gives a net
torque, and tends to fill the gap.


\section{Gap profiles}
\label{sec:Semi-analytic_results}
\label{sub:semi-analytic_formula}

In the last section, the pressure torque has been numerically computed
in different cases. It has been shown that, { when the disk is in}
equilibrium, the pressure, gravity and viscous torques cancel
out. This suggests that it should be possible to compute a priori the
shape of the gap by imposing that this equilibrium \eqref{eq:sum=0} is
respected. Indeed, the viscous and pressure torques depend on the
relative radial gradient of the azimuth-averaged density, whereas the
gravity torque { has no direct dependence on it.} Therefore, on a
given trajectory, there must be a value of this gradient that
corresponds to the exact equilibrium between these three torques.

{ Our} semi-numerical algorithm for the computation of this
equilibrium value is described in appendix~B. The results are shown in
Fig.~\ref{fig:semi-anal-sol} (crosses) and satisfactorily agree with
the real values measured in the corresponding numerical simulation
(solid curve), {\it i.e.} the simulation from which the streamlines
used by the algorithm have been obtained.

The knowledge of the relative radial gradient of the azimuth-averaged
density as a function of the radial distance enables us to construct a
gap profile by simple numerical step by step integration, starting
from a boundary condition. In the secondary panel of
Fig.~\ref{fig:semi-anal-sol} this integrated profile (dashed curve) is
plotted against the real one from the considered simulation. The match
between the two profiles is almost perfect, which again proves that
the gap profile is set by the balance between the three torques due to
gravity, viscosity, and pressure.

\begin{figure}[t!]
\includegraphics[width=0.7\linewidth,angle=270]{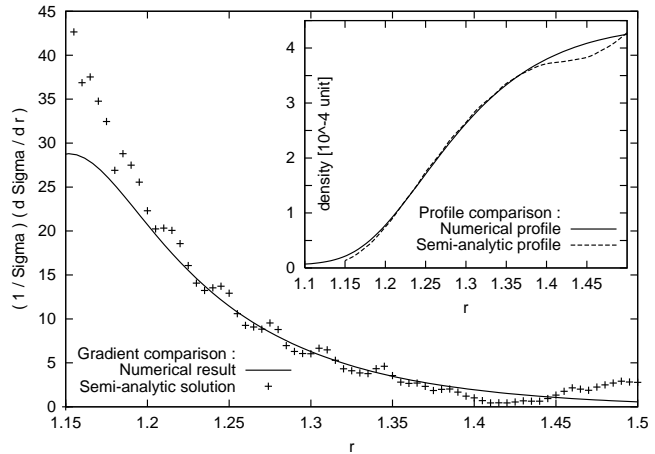}
\caption{\small The crosses show the relative radial gradient of the
surface density as computed by the algorithm described in
Appendix~B. The solid curve shows the same quantity, measured from the
numerical simulation from which the streamlines used in the algorithm
have been taken (aspect ratio = $0.05$, viscosity = $10^{-5.5}$,
planet mass = $10^{- 3}$). From the algorithm, the gap
profile is computed (dashed curve in the little box), and compared to
that obtained in the numerical simulation (solid curve). }
\label{fig:semi-anal-sol}
\end{figure}

\subsection{An explicit equation for the gap profile}

We now wish to go beyond the semi-numerical algorithm of Appendix~B
and obtain an approximate analytic expression for the pressure torque,
to be used in an explicit differential equation for the gap profile.

As we have seen above, for a given streamline, the absolute value of
the pressure torque is an increasing function of the relative radial
gradient of the azimuthally averaged surface density. Furthermore, in
a disk with no density gradient, the pressure torque must be
zero. Thus, we approximate the dependence of the pressure torque on
the relative radial density gradient with a linear function\,:
\begin{equation*}
t_P= - a(r)\, \left(\frac{{\rm d}{\Sigma}}{\Sigma\ {\rm d}r}\right)\ .
\end{equation*}

Before looking for a numerical approximation of the function $a(r)$,
we { make two} considerations on its functional dependence on the
scale height of the disk and on the mass of the planet.

{ First,} because of Eq.~\eqref{eq:tPtube}, $a(r)$ is necessarily
proportional to ${c_s}^2$. As ${c_s}$ is proportionnal to {the scale
height $H$}, we can write $a=(H/r)^2 a'(r)$.

{ Second, in the limit of} negligible viscosity, scaling the aspect
ratio $H/r$ proportionally to $R_H/r_p$, and adopting $R_H$ as basic
unit of length, the equation of motion becomes independant of the
planet mass (Korycansky and Papaloizou, 1996). Thus, if the disk
aspect ratio scales with the planet Hill radius, the resulting surface
density ${\Sigma}$ at equilibrium is a function of $\Delta/R_H$ only.
Consequently ${\rm d}{\Sigma}/(\Sigma {\rm d}r)$ is a function of
$\Delta/R_H$, divided by $R_H$. As the gravity torque $t_g$ is
proportional to ${R_H}^{2}r(\Delta/R_H)^{-4}$ (see \eqref{eq:t_g} ),
the equilibrium $t_g=t_p$ can hold if and only if $a'(r)=rR_H\
a''(\Delta/R_H)\ r_p{\Omega_p}^2$, where $a''(\Delta/R_H)$ is a
dimensionless function and the constant factor $r_p{\Omega_p}^2$
stands for homogeneity reasons.

To evaluate the function $a''$, we use numerical
simulations from which we measure the pressure torque and the relative
radial gradient of the surface density. In practice, we consider two
simulations\,: (i) the reference one, with a Jupiter mass planet in a
disk with aspect ratio = $0.05$ and viscosity = $10^{-5.5}$, which
gives information for $\Delta/R_H$ in the range 2--7, and (ii) a
similar simulation but with viscosity $\nu=10^{-6.5}$ which, because
of its wider gap, allows us to better estimate the asymptotic behavior
of $a''$ at large $\Delta$. We find that $a''(\Delta/R_H)$ can be
{approximately fitted} by the function
\begin{equation}
a''\left(\frac{\Delta}{R_H}\right)=\frac{1}{8}\left|\frac{\Delta}{R_H}\right|^{-1.2}+
 200 \left|\frac{\Delta}{R_H}\right|^{-10}\ .
\label{eq:asecond}
\end{equation}

Equation \eqref{eq:asecond} has been determined for the external part
of the disk ($\Delta>0$), outside of the horseshoe region. However,
assuming that the streamlines are symmetric relative to the position
of the planet, the same expression can be applied in the inner part of
the disk, which justifies the absolute value of $\Delta$. In fact, to
represent the inner edge of the gap, just rotate
Fig.~\ref{fig:gradient} by 180 degrees, and it becomes evident that
a negative density gradient leads to a positive torque.

Note that Eq.~\eqref{eq:asecond} has been determined with
reference to the streamlines corresponding to the case with
$q=10^{-3}$, $\nu=10^{-5.5}$, and $H/r=5\%$. However, the exact shape
of the streamlines depends on $q$, $\nu$, and $H/r$, even in rescaled
coordinates. {We neglect this dependence at this stage.} 

Thus, we assume that \eqref{eq:asecond} is
valid for { any} value of $\nu$ and $H/r$ and $a''$ depends on $q$ only
via $R_H$. This approximation has the advantage of
providing us an analytic expression for the computation of the gap
profiles. In fact, the disk equilibrium equation \eqref{eq:sum=0}
becomes\,:
\begin{equation}
\left(\frac{R_H}{{\Sigma}}\frac{{\rm d}{\Sigma}}{{\rm d}r}\right) =
\frac{t_g - \frac{3}{4}\nu\Omega}
{\left(\frac{H}{r}\right)^2 r\,r_p{\Omega_p}^{\!2} a''+\frac{3}{2}\nu \frac{r}{R_H}\Omega}
\label{eq:gap_analytic}
\end{equation}
with $a''$ and $t_g$ given in Eq.~\eqref{eq:asecond} and \eqref{eq:t_g}
respectively.

The right hand side of Eq.~\eqref{eq:gap_analytic} is independent of
${\Sigma}$ and is an explicit function of $r$. This differential
equation can be integrated, once a boundary condition
${\Sigma}(r_0)$ is given. Unfortunately the integral has no
analytic solution, so that it has to be computed numerically.

\begin{figure}[t!]
\includegraphics[width=1.\linewidth]{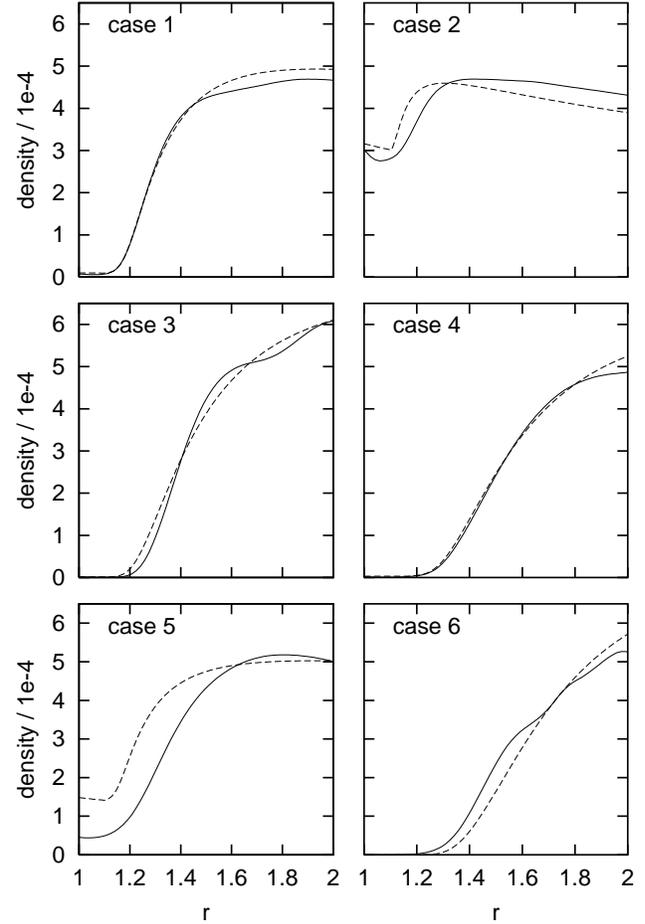}
\caption{\small Comparison of numerical results (plain lines) and
analytic profiles given by Eq.~\eqref{eq:gap_analytic} (dashed lines).
{\it Case 1\,:} reference case\,: $q=10^{-3}$, $H/r=5\%$, $\nu=10^{-5.5}$.
{\it Case 2\,:} more viscous case\,: $q=10^{-3}$, $H/r=5\%$, $\nu=10^{-4}$.
{\it Case 3\,:} less viscous case\,: $q=10^{-3}$, $H/r=5\%$, $\nu=10^{-6.5}$.
{\it Case 4\,:} scaled case\,: $q=3.10^{-3}$, $H/r=7.2\%$, $\nu=10^{-5.5}$.
{\it Case 5\,:} thicker disk case\,: $q=10^{-3}$, $H/r=10\%$, $\nu=10^{-5.5}$.
{\it Case 6\,:} more massive case\,: $q=3.10^{-3}$, $H/r=5\%$, $\nu=10^{-5.5}$.
}
\label{fig:semi-anal_vs_num}
\end{figure}

Figure~\ref{fig:semi-anal_vs_num} shows comparisons of the gap
profiles obtained in numerical simulations with those obtained with
the integration of Eq.~\eqref{eq:gap_analytic}, for six different
cases with different viscosities or aspect ratios and planetary mass
(see figure caption for a list of parameters). { The comparisons
are done only for the outer part of the disk, because in the inner
part, the effect of the boundary condition, not considered in our
model, is too prominent in the numerical results.}  In the integration
of Eq.~\eqref{eq:gap_analytic}{, ${\Sigma}$ has been set}
equal to the value found in the numerical simulations at a point on
the brink of the gap, so as to allow an easier comparison between the
numerical and semi-analytic density gradients at the edge of the
gap. We remark that in case~1, the semi-analytic gap profile
matches almost perfectly the numerical profile. This is not
surprising, because this is the reference case for which the
streamlines have been computed, so that our expression
\eqref{eq:asecond} is virtually exact.

In cases~2 and 3, we change the viscosity and keep the same planet
mass and aspect ratio as in case~1. Now, the agreement between the
numerical and semi-analytic profiles is less good. In particular, in
the high-viscosity case, the real density gradient is shallower than
the one we compute, while in the low-viscosity case it is
steeper. This is because the real streamlines are not identical to
those for which Eq.~\eqref{eq:asecond} has been computed. In the more
viscous case, the equilibrium in the disk is achieved with a weaker
pressure torque. The distortion of the streamlines at the wake is
dictated by the difference between the local gravity and pressure
torques. Thus, a weaker pressure torque gives streamlines that are
more distorted at the wake than in our reference case. But, as
sketched in Fig.~\ref{fig:gradient}, the more a streamline is
distorted, the more efficient it is in producing a pressure torque
from a radial surface density gradient. Consequently the pressure
torque required to set the equilibrium in the disk is achieved with a
smaller density gradient than the one needed if the streamlines were
as in the reference case. The opposite holds in the less viscous case.

{ In case~4, we increase the mass of the planet and the disk aspect
ratio, in a way such that $H/R_H$ is the same as in case~1. The
viscosity is also the same as in case~1, and the agreement between the
model and the simulation is equally good.

Finally, in case~5 and 6, we change $H/R_H$. In case~5, we keep the
planet mass and viscosity of case~1 but increase the aspect ratio to
$10\%$\,; the model gap is quite narrower and shallower than the
numerical one. In case~6, we use the same planet as in case~4 but with
the aspect ratio and viscosity of case~1, which gives as smaller
$H/R_H$\,; the gap that our model predicts is now slightly wider than
the one obtained in the numerical calculation. The interpretation for
the disagreements observed in cases 5 and 6 is the same as that
offered for cases 2 and 3.}

\subsection{Note on disk evolution during gap opening}
\label{sub:bump}

Figure~\ref{fig:semi-anal_vs_num} shows significant differences in the
value of $\Sigma$ between the numerical simulation and the analytic
expression. However, we stress that a large difference in $\Sigma$ can
correspond to almost no difference in { the relative slope}
$(\frac{1}{\Sigma} \frac{{\rm d}\Sigma}{{\rm d}r})$. For instance, in
the case with viscosity equal to $10^{-4}$ (case 2), the surface
density profiles for $r>1.6$ seem quite different, but in fact, they
have the same relative slope.

\begin{figure}[t!]
\includegraphics[width=0.7\linewidth,angle=270]{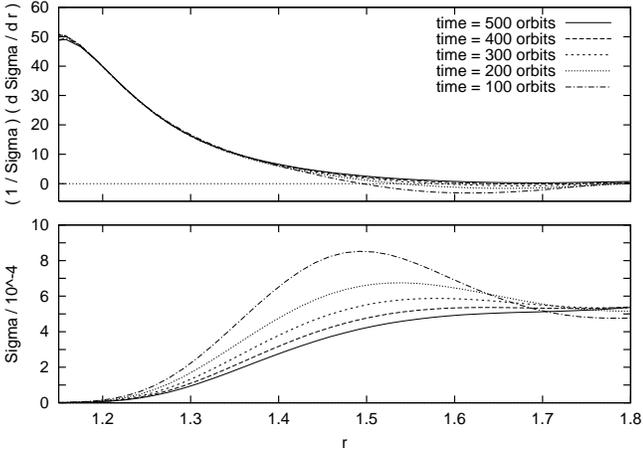}
\caption{\small Evolution with time of the profile of the external edge
of the gap opened by a Jupiter mass planet in a $5\%$ aspect ratio
disk, with a $10^{-6.5}$ viscosity. Top panel shows the relative
radial gradient of the density\,; the fact that the curves overlap
argues that the equilibrium function has been rapidly reached. Bottom
panel shows the evolution of the surface density at the same times\,;
a 'bump' appears, { which is a} consequence of the matter removed
from the gap, and it is eroded on a viscous time scale. This happens
without modifying substantially the relative slope.}
\label{fig:bump-evol}
\end{figure}

As we have shown above, it is the relative slope $(\frac{1}{\Sigma}
\frac{{\rm d}\Sigma}{{\rm d}r})$ that sets the equilibrium. This
equilibrium must be reached quickly, on a time scale independent of
viscosity. In fact, in absence of equilibrium, the fluid elements are
displaced radially over a synodic period and the trajectories are not
periodic. This corresponds to the opening of the gap. Then, once the
equilibrium is almost set, the value of $\Sigma$ can still
significantly evolve on a long (viscous) time scale, but keeping
$(\frac{1}{\Sigma}\frac{{\rm d}\Sigma}{{\rm d}r})$ essentially
unchanged. This fact is illustrated on Fig.~\ref{fig:bump-evol}, which
compares the evolution of $(\frac{1}{\Sigma} \frac{{\rm d}\Sigma}{{\rm
d}r})$ (top panel) with the evolution of $\Sigma$ (bottom panel) for a
weakly viscous case ($\nu=10^{-6.5}$). { This behavior explains
why, when simulating the gap opening in low viscosity disks, the
surface density profile seems to have attained a stationary solution
within a limited number of planetary orbits, despite that the presence a
prominent 'bump' at the outer edge of the gap indicates that there is
still room for evolution (see Fig.~\ref{fig:figure1poster}).}

{In the inner disk, once the gap profile is set in terms of
relative slope, we expect that the density $\Sigma$ decreases
on a viscous timescale, because of the accretion on the central
star. In the approximation of a fixed planet, this viscous evolution
would lead to the formation of an inner hole in the disk, extended
up to the planet position.}


\section{Dependence of gap profiles on viscosity and aspect ratio}
\label{sec:gap_profil_visc_AR}

In the previous section, we have presented a semi-analytic method to
compute gap profiles. It gave overall satisfactory results, as shown
in Fig.~\ref{fig:semi-anal_vs_num}. In this section, we use our
method to explore the dependence of the gap profile on the two key
parameters of the disk\,: viscosity $\nu$ and aspect ratio $H/r$. In
particular, we wish to revisit, with a unitary approach, the gap
opening criteria mentioned in
section~\ref{sec:Gravity_and_Viscous_Torques}\,:

\noindent (i) the viscosity needs to be smaller than a threshold value. 
According to Eq.~\eqref{eq:nucrit}, in our case of a Jupiter
mass planet this value is
$\nu_{\rm crit}\approx 10^{-4}$.

\noindent (ii) The disk height at the location of the planet needs to
be smaller than $\sim R_H$. For a Jupiter mass planet it corresponds
to an aspect ratio $\sim 7\%$.

\begin{figure}[t!]
\includegraphics[width=0.7\linewidth,angle=270]{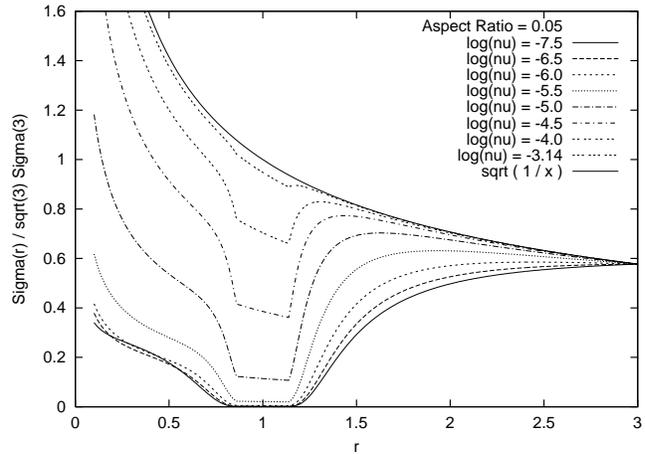}
\caption{\small Analytical gap profiles given by
Eq.~\eqref{eq:gap_analytic} for different viscosities. The gap deepens
as viscosity decreases, but its width remains bounded, even for
$\nu=0$.}
\label{fig:Pa-v}
\end{figure}

\begin{figure}[t!]
\includegraphics[width=0.7\linewidth,angle=270]{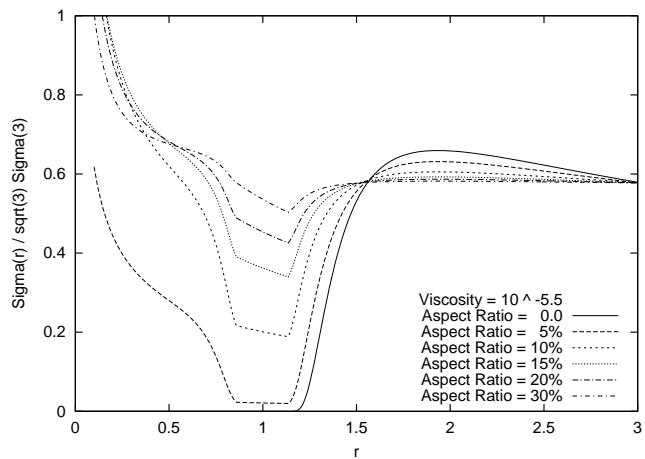}
\caption{\small Analytical gap profiles given by
Eq.~\eqref{eq:gap_analytic} for different aspect ratios. The gap
deepens as the aspect ratio decreases.}
\label{fig:Pa-AR}
\end{figure}

{ In the computation of the gap profiles by integration of
Eq.~\eqref{eq:gap_analytic}, two problems are encountered.

First, a boundary condition is needed. This choice is arbitrary, but
in principle it should be consistent with the steady state of the
disk. However, the steady state is an academic concept which exists
only if the density is kept fixed somewhere in the disk, otherwise the
disk spreads to infinity following Lynden-Bell \& Pringle (1974)
equation. In our numerical code, the surface density is kept equal to
the unperturbed value at the outer boundary of the grid ($r=3$). Thus,
for the solutions of Eq.~\eqref{eq:gap_analytic} presented in
Figs.~\ref{fig:Pa-v} and \ref{fig:Pa-AR}, we impose
$\Sigma(r=3)=1/\sqrt{3}$. In this way, our solution should correspond
to the steady state solution that the code would converge
to. Moreover, this choice allows a direct comparison of our profiles
with those obtained with Varni\`ere {\it et al.} model, illustrated in
the top panel of Fig.~\ref{fig:figure1poster} using the same boundary
condition.

The steady state solution provided by the numerical simulation does
not depend on the size of the grid, provided that the boundaries are
sufficiently far from the planet (negligible differential planetary
and pressure torques, or equivalently, negligible wave carried angular
momentum flux, as in our nominal case --\,see Appendix~C ). This
required size increases with decreasing viscosity because the radial
range over which the wave is damped increases.

Thus, in a very low viscosity case, the steady state solution obtained
by the numerical simulation over an extended disk would be different
from the model profiles given on Fig.~\ref{fig:Pa-v}. However, our
model profiles would bound the gap observed in the numerical
simulation as long as the normalized surface density at $r=3$ does not
decrease below $1/\sqrt{3}$. In such low viscosity cases, this happens
after an exceedingly long time. Thus, we claim that our model profiles
are significant for the description of gaps in realistic disks.
}

The second problem concerns the treatment of the horseshoe region. The
gravity and pressure torques, $t_g$ and $t_P$, are considered null in
the horseshoe region $|\Delta|<2\,R_H$. The depth of the gap is thus
set by the value of the density at $r_P+2\,R_H$. At the edges of the
gap, the slope is very steep, so that a little change in the assumed
width of the horseshoe region leads to a major change in the gap
depth. This is a limitation of our results from a quantitative point
of view. Though, it does not change the qualitative evolution of the
gap profiles with respect to the disk parameters.

This sensitivity {to} the width of the horseshoe region is also a
problem for the construction of the surface density profile in the
inner disk. The integration for the inner disk starts from
$r_P-2\,R_H$ down to $r=0$, with the density at the bottom of the gap
acting as { the} boundary condition. In principle, if the gap
profile is symmetric, the errors at the right hand side and left hand
side borders of the gap compensate each other\,: the value of the
surface density at the bottom of the gap is not quantitatively
correct, but the density profile in the inner disk is realistic.

Figure~\ref{fig:Pa-v} shows the results of our semi-analytic
calculation for a fixed value of the aspect ratio ($5\%$) and
different values of the viscosity $\nu$ (from $0$ to $10^{-3}$
in normalized units). Figure~\ref{fig:Pa-AR} keeps the viscosity
$\nu=10^{-5.5}$ and explores the dependence of the gap profile on the
disk aspect ratio (from $0$ to $30\%$). The plotted curves naturally
order themselves from bottom to top, from the less viscous case
(respectively the smallest aspect ratio) to the most viscous case
(respectively the biggest aspect ratio). Notice that this progression
does not represent an evolution with time, but different steady state
gap profiles, for different parameters.

\paragraph{dependence on viscosity\,:}
First of all, we remark on Fig.~\ref{fig:Pa-v} that the different
shapes of the gaps qualitatively agree with those computed with
numerical simulations, shown in the bottom panel of
Fig.~\ref{fig:figure1poster}. Indeed, not only do we get deeper and
wider gaps as viscosity decreases, but we also correctly reproduce the
limited gap width achieved { in the small viscosity cases}. This
means we have solved the problem that {initially} motivated our
investigation.

\noindent
As viscosity increases, the gap is filled with gas, and the profile
tends to the unperturbed profile set by the sole viscous effects\,:
$\Sigma\propto 1/\sqrt{r}$ (see section
\ref{sec:Gravity_and_Viscous_Torques}). Nevertheless, it is hard to
determine a precise threshold value for gap opening, for at least two
reasons. The first one is that the gap profiles have a smooth
dependence on the viscosity. The concept of threshold viscosity for
gap opening does not hold. The gap gradually increases in depth over a
range of viscosities { of about one order of magnitude.} The second
reason is that the depth of our gaps is very sensitive { to} the
assumed width of the horseshoe region, as discussed above. Thus, there
is some uncertainty on the value of the viscosity that makes the gap
become only a dip. { Assuming that a gap is opened if the surface
density falls below $10\%$ of the unperturbed density, we find that
$\nu_{\rm crit}\approx 10^{-5}$.  We remind that the `classical'
threshold for gap opening is $\nu_{\rm crit}\approx10^{-4}$. However,
the numerical experiments in Fig.~\ref{fig:figure1poster} (bottom
panel) also suggest that $\Sigma\sim 0.1$ at the bottom of the gap is
achieved for $\nu\sim 10^{-5}$.}

\paragraph{dependence on aspect ratio\,:}
Consider now Fig.~\ref{fig:Pa-AR}. We see a smooth evolution from deep
gaps to shallow or inexistent gaps with increasing aspect ratios. This
is easy to understand, because $(H/r)^2$ is a multiplicative
coefficient in the expression of the pressure torque (see
section~\ref{sec:Semi-analytic_results}). Therefore, the larger
$(H/r)$, the shallower needs to be the relative slope at the edge of
the gap to achieve the equilibrium \eqref{eq:sum=0}. As in the
previous case, it is not possible to determine a threshold value of
$(H/r)$ for gap opening, but we find that the `classical' value
$(H/r)_{\rm crit}\approx 0.07$ { corresponds to about $90\%$
depletion in the gap}.

\vskip 15pt

More generally, with our approach we find that the viscosity required
to fill the gap is a decreasing function of the aspect ratio. If the
aspect ratio is too large, the gap cannot be opened whatever the
viscosity. To our knowledge, this is the first time that an analytic
approach gives the correct description of the evolution of the gap
profile with respect to {\it both} disk viscosity {\it and} aspect
ratio.


\section{A new generalized criterion for gap opening}
\label{sec:new_criterion}

To go beyond the qualitative considerations of the previous section,
we try to generalize the gap opening criterion with an expression that
involves simultaneously the three main parameters of the problem\,:
mass of the planet, scale height of the disk and viscosity.

We start with a few considerations on two limiting cases. In the
zero viscosity limit, as we have seen in section
\ref{sec:Semi-analytic_results}, changing the scale height of the disk
{ in proportion} to the Hill radius of the planet preserves the gap
profile in scaled units $\Delta/R_H$. This means that, whatever depth
threshold is adopted for the definition of `gap', { the threshold
value of $H$ for gap opening in the zero viscosity limit, $H_0$, is
proportional to $R_H$\,:
$$H_0 \propto R_H \propto q^{\frac{1}{3}}\ .$$}

In the infinitely thin disk limit ($H/r\to 0$), the disk equilibrium
is set by the equation $t_g=t_\nu$. At the border of the gap where the
slope of the surface density is relevant, $t_\nu$ is proportional to
$\nu \frac{r\,{\rm d}\Sigma}{\Sigma\,{\rm d}r}$. By changing the mass
of the planet, the gravity torque changes proportionally to
${R_H}^{2}r(\Delta/R_H)^{-4}$. If the viscosity $\nu$ is changed
proportionally to ${R_H}^{3}\propto q$, then the surface density
profile $\Sigma$ remains an invariant function of $\Delta/R_H$.  Thus,
independently of the adopted definition of `gap' as in the previous
case, { the threshold viscosity for gap opening in the infinitely
thin disk limit, $\nu_0$,} scales proportionally to $q$. This is
consistent with the gap opening criterion given in Bryden {\it et al.}
(1999).
$$\nu_0 \propto q$$

We now come to the general case where neither $H$ nor $\nu$ are null.
From the considerations above and Eq.~\eqref{eq:gap_analytic} it
is evident that a change in the planet mass $q$ can give an invariant
surface density profile in scaled units $\Delta/R_H$ provided that $H$
is changed proportionally to $q^{\frac{1}{3}}$ {\it and} $\nu$ is
changed proportionally to $q$.

\begin{figure}[t!]
\includegraphics[width=0.7\linewidth,angle=270]{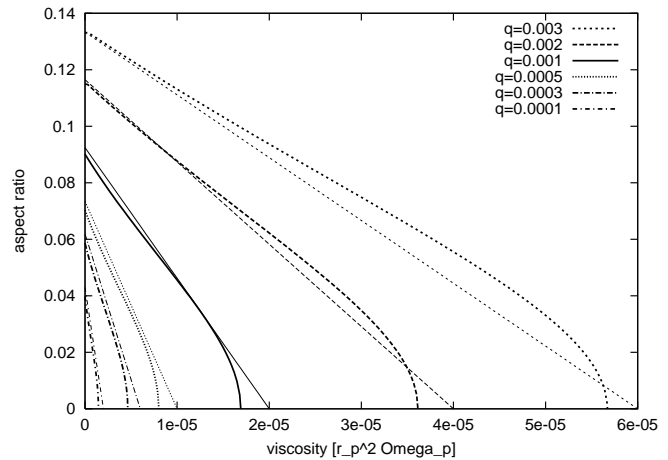}
\caption{\small The bold curves represent the values of $(H/r)$
that make the gap depth be $10\%$ of the unperturbed density for given
values of $\nu$. They have been computed from the solutions of
Eq. \eqref{eq:gap_analytic}. Each curve corresponds to a planet
mass. The thin lines represent our linear approximations given by
Eq. \eqref{Crida_function} for the corresponding planet mass.}
\label{fig:thresholds}
\end{figure}

The most complicated case that remains to be analyzed is that where
$q$ is constant, but $H$ and $\nu$ are changed. It is evident from
Eq.~\eqref{eq:gap_analytic} that it is not possible to have an
invariant surface density profile by decreasing $H$ and increasing
$\nu$ or vice-versa. The question is therefore how to keep { the
central gap depth invariant, despite changes in the gap profile}. We
answer this question using our semi-analytic calculation of gap
profiles, based on the integration of Eq.~\eqref{eq:gap_analytic}. For
this, we define --\,arbitrarily\,-- that the minimal depth that
defines a gap is 1/10 of the unperturbed disk density at $r=r_p$.
Figure~\ref{fig:thresholds} shows as bold lines, for six different
values of the planet mass, the relationships $H$ vs. $\nu$ that
preserve such central gap depth.

As one can see, these relationships are almost linear.

{ We can fit each one with a relation of type $H/H_0 + \nu/\nu_0
=1$, where $H_0$ and $\nu_0$ have been defined above. As we have
$\nu_0\propto q$ and $H_0\propto q^{1/3}$, we can derive
a general relation involving $H$, $\nu$ and $q$ that approximately
describes all the curves plotted on Fig.~\ref{fig:thresholds}, and
thus a general criterion for gap opening. Denoting by ${\cal R}$ the
Reynolds number ${r_p}^{\!2}\Omega_p/\nu$, we find that a gap is
opened if $q$, $H$ and ${\cal R}$ satisfy the following inequality\,:

\begin{equation}
\frac{3}{4}\frac{H}{R_H}+\frac{50}{q\mathcal{R}} \lesssim 1\ .
\label{Crida_function}
\end{equation}

The thin lines on Fig.~\ref{fig:thresholds} correspond to the limit
case $\frac{3}{4}\frac{H}{R_H}+\frac{50}{q\mathcal{R}}=1$, for any of
the six considered values of $q$.

}


\section{Conclusion}

In this paper, we have analyzed in detail the process of gap opening
in proto-planetary disks. In this respect, a key problem is to
calculate which fraction of the torque exerted by the planet is
locally deposited in the disk and which fraction is transported away
by pressure supported waves. We have shown that the angular momentum
evacuated by the waves can be computed as a pressure torque. We found
that the steady state of the disk is set by the equilibrium among the
total gravity torque, the viscous torque, and the pressure
torque. From this consideration, we have built a semi-analytical
algorithm that, given viscosity and aspect ratio, provides the
equilibrium profile of the surface density of the disk, { enabling}
us to explore the gap shape for a large range of parameters.

This work { has} two types of application. It can be used to
achieve a first realistic estimate of the width and depth of gaps in
various situations, in view of the future high resolution observations
of proto-planetary disks (with ALMA or the SKA projects). It can also
give equilibrium gap profiles to be used as a starting condition in
numerical simulations if one wants to avoid the intermediate,
cpu-consuming phase which leads to the steady state.

Our work is not fully analytic. Indeed our final equation
\eqref{eq:gap_analytic} involves a function $a''$ which we
approximated by the ansatz function~\eqref{eq:asecond}, with
coefficients determined with respect to a reference numerical
simulation. Also the gravity torque \eqref{eq:t_g} has been refined
using fits to the reference numerical results. As a consequence, if
our model matches the results of the reference numerical simulation,
it { still is in satisfactory agreement} with the results of other
numerical simulations.

Moreover, the equilibrium profile that we obtain corresponds to the
equilibrium configuration of the disk at infinite time in presence of
a non-migrating planet, which is evidently an ideal case.  { Our
model is two-dimensional, intended to approximate the behavior of a
vertically isothermal 3D disk\,; in a 3D, thermally stratified disk,
the density waves would not propagate exactly the same way (Bate {\it
et al.}, 2003) and consequently the pressure torque is expected to be
somewhat different. Finally, we have assumed a constant kinematic
viscosity\,; in reality, in the regions where the perturbations are
nonlinear, the effective viscosity depends on the local planet's
gravitational torque (Goodman and Rafikov, 2001), although this
dependance may be weak (Papaloizou {\it et al.}, 2004). }

In spite of these limitations, our work { clearly demonstrates}
the fundamental role of the pressure in setting the equilibrium of the
disk. Moreover, it gives a correct, nearly quantitative, description
of the evolution of the gap profile with respect to the key parameters
of the problem\,: planet mass, viscosity and aspect ratio. From this
we derive a new general criterion for gap opening, involving
simultaneously these three parameters.

Our work shows why the width of the gap is bounded even in the case
with very small viscosity, which was the open problem that originally
motivated our work. It provides a conceptual unification of the two
classically, but independently derived, criteria for gap opening,
based on threshold viscosity and aspect ratio.

{ {\sc Aknowledgements\,:} The authors thank anonymous referees for
their carefull reading and their interesting remarks. We also thank
David O'Brien for his comments on the manuscript.}


\section{Appendix}

\subsection*{A: trajectories and streamlines computation.}
\label{sub:traj_stream}

{{ In a} steady state, trajectories and streamlines
coincide. Computing the streamlines is then equivalent to computing
the trajectories.} In principle, to compute a trajectory it is enough
to integrate the velocity field. The latter is defined on the grid and
output by the code, from which the velocity at any point of the disk
can be computed by interpolation. However, using this procedure, the
resulting trajectories would in general not be periodic, as a
consequence of the accumulation of the integration and interpolation
errors. This is a serious problem, because the loss of periodicity
would introduce a spurious change of angular momentum, namely a
spurious torque.

To obtain perfectly periodic streamlines, we used the following
algorithm, that for simplicity we detail for the outer part of the
disk ($r>r_p$). We first compute a trajectory from
$(r=r_0,\theta=\pi)$ to $\theta=-\pi$ by simple numerical integration
of the velocity field (the integration runs from $\pi$ to $-\pi$
because $r_0>r_p$, so that the fluid element rotates clockwise in the
corotating frame). This gives a first curve $r^{(1)}(\theta)$, defined
on the interval $[-\pi,\pi]$. { By definition} $r^{(1)}(\pi)=r_0$,
but $r^{(1)}(-\pi)$ is in general close but not equal to $r_0$,
because of numerical errors, as said above. On this trajectory, we
calculate $\frac{v_r}{v_\theta}(r^{(1)}(\theta),\theta) \equiv
f^{(1)}(\theta)$. This is a pseudo-derivative of $r^{(1)}$, {
i. e.} the slope of the tangent to the curve according to the velocity
field. It should be equal to ${\rm d}r^{(1)}/{\rm d}\theta$, but {
is} not exactly { equal} because of the numerical errors in the
computation of $r^{(1)}$. Then, we compute the Fourier coefficients
$f^{(1)}_n$ of $f^{(1)}(\theta)$. The first one $f^{(1)}_0$ is real,
and corresponds to the mean of $f^{(1)}$, namely to a radial drift. It
is not zero as $r^{(1)}$ is not { exactly} periodic{, and
therefore we set it to zero}. The pseudo-derivative of $r^{(1)}$ with
respect to $\theta$ is thus modified. To get back to a trajectory, we
integrate this modified pseudo-derivative. We denote the new
trajectory by $r^{(2)}(\theta)$. This trajectory is periodic by
construction as its zeroth order Fourier coefficient is null. From
$r^{(2)}$, we repeat the algorithm to find $r^{(3)}$, and so on, until
the algorithm converges to a { fixed} point. This { fixed} point
$r(\theta)$ is a periodic trajectory by construction. It fits the
velocity field, as it verifies $\frac{{\rm d}\,r(\theta)}{{\rm
d}\theta} = \frac{v_r}{v_\theta} (r(\theta),\theta)$, provided that
the zeroth order Fourier coefficient of its pseudo-derivative is
negligible. If it isn't, it means that the algorithm failed. This
happens in particular if the real streamlines are not periodic because
a steady state has not been reached yet.

In practice, for the implementation of this algorithm, we used
simulations computed over a grid with a larger resolution than that
used in section~\ref{sec:Gravity_and_Viscous_Torques}. More precisely,
we have used 512 cells in radius and 1024 in azimuth. The number of
points used to compute the Fourier coefficients of the
pseudo-derivative was 1024 too. In all cases, the algorithm explained
above converged, and the zeroth order Fourier coefficient of the final
pseudo-derivative was negligible (less than $10^{-3}$ in our
normalized units, even $10^{-4}$ for all trajectories with
$r(\pi)>1.2$).

\subsection*{B: a semi-numerical algorithm for the calculation of the equilibrium surface density slope.}
\label{sub:semi-analytic_profiles}

We present an algorithm that, given the shape of the streamlines,
computes the relative surface density radial gradient that ensures the
equilibrium condition \eqref{eq:sum=0}. This is done in two
steps. First, we design a procedure that evaluates $t_P$ on each
streamline, for any given value of $\frac{1}{\Sigma}\frac{{\rm
d}\Sigma}{{\rm d}r}$. Second, we solve numerically the implicit
equation { for} $\frac{1}{\Sigma}\frac{{\rm d}\Sigma}{{\rm d}r}$
given by Eq.~\eqref{eq:sum=0}.

\paragraph{First step\,: computation of the torques}
The streamlines are ordered with respect to increasing distance to the
central star, so that $r_i(\theta)<r_{i+1}(\theta)$ for every $i$,
$\theta$. { We call the $i$th {\it streamtube} the zone around the
$i$th streamline\,: $\{\,(r_{i-1}+r_i)/2 < r < (r_i+r_{i+1})/2\,\}$.}
A total mass $m_i$ or mean density ${\Sigma}_i$ can be
imposed to be carried by a given { streamtube} $i$. Because the steady
state is reached, the flux of matter { in streamtube $i$} is
constant with respect to time and azimuth, and is $F_i=m_i/T_i$, where
$T_i$ is the synodic period along the streamline. Thus, the mass has
to be distributed in the streamtube in such a way that the flux
\begin{equation}
F(\theta)=\Sigma(r_i(\theta),\theta)
v_\theta(r_i(\theta),\theta)[r_{i+1}(\theta)-r_{i-1}(\theta)]/2
\label{F}
\end{equation}
is equal to $F_i$ for all $\theta$. The azimuthal speed
$v_\theta(r_i(\theta),\theta)$ can be obtained by interpolation from
the output of the numerical code\,; the local density
$\Sigma(r_i(\theta),\theta)$ is therefore the only unknown in
Eq.~\eqref{F}, so that one has\,:
\begin{equation}
\Sigma(r_i(\theta),\theta) = 2F_i/v_\theta(r_i(\theta),\theta)
\left[r_{i+1}(\theta)-r_{i-1}(\theta)\right]\ .
\label{Sigma}
\end{equation}

Any relative radial density gradient $(1/{\Sigma}) ({\rm
d}{\Sigma}/{\rm d}r)$ around the $i$th streamline can be created by
imposing appropriate values for ${\Sigma}_{i+1}$ and
${\Sigma}_{i-1}$. The masses $m_{i+1}$ and $m_{i-1}$ carried by
the streamlines are obtained by multiplying the mean surface densities
by the areas of the stream tubes. Then, the local densities
are computed using Eq.~\eqref{Sigma}.

Once the streamlines and the local densities are known, the numerical
computation of the pressure torque can be done using
Eq.~\eqref{eq:good_torques_tP}. The partial derivative of the density with
respect to the azimuth is delicate to compute. Indeed, from
Eq.~\eqref{Sigma} we know $\Sigma(r_i(\theta),\theta)$ only on a discrete
set of values $r_i(\theta)$. To compute $(\partial \Sigma /
\partial\theta)$ at the location $(r_i(\theta),\theta)$ we need to
know $\Sigma(r_i(\theta),\theta\pm \delta\theta)$, for some small
$\delta\theta$. This is computed by interpolation between
$\Sigma(r_j(\theta\pm \delta\theta),\theta\pm \delta\theta)$ and
$\Sigma(r_{j+1}(\theta\pm \delta\theta),\theta\pm \delta\theta)$, where the
$j$th streamline is chosen such that $r_j(\theta\pm \delta\theta)<
r_i(\theta) < r_{j+1}(\theta\pm \delta\theta)$.

The viscous and gravity torques are given by expressions
\eqref{eq:dTnu} and \eqref{eq:t_g} respectively, with $r\equiv
r(\pi)$. { Expression~\eqref{eq:dTnu} is preferred to expression
\eqref{eq:good_torques_tnu}, despite of its limitations in the very
vicinity of the planet (see Fig.~\ref{fig:TpTgTnu_r0}) because it is
simple and explicit.}

\paragraph{Second step\,: computation of the gap profile}
To obtain the density profile we impose that the sum of the three
torques vanishes on every streamline. Thus, for each streamline, the
goal is to find the value of the { relative surface density slope}
$s=(\frac{1}{{\Sigma}}\frac{{\rm d}{\Sigma}}{{\rm d}r})$ that makes
the total torque $t_{\rm total}(s)=\left(t_P+t_\nu+t_g\right)$ equal
to zero. As the pressure torque is numerically computed, the solution
can be found only numerically. We use a secant method algorithm,
described next.

A first value $s_0$ of $s$ is arbitrarily chosen (typically $0$ or the
solution found on a neighboring streamline). Then, another value $s_1$
is taken (for instance $s_0+100\,t_{\rm total}(s_0)$). The secant
method algorithm { is then used}. The value chosen for $s_2$ is\,:
$s_1-t_{\rm total}(s_1).(s_1-s_0)/\left[t_{\rm total}(s_1)-t_{\rm
total}(s_0)\right]$. A sequence $(s_n)_{n=0,1,\ldots}$ is build this
way. At each step, the tested value is\,: $s_n=s_{n-1}-t_{\rm
total}(s_{n-1}).(s_{n-1}-s_{n-2}) / \left[t_{\rm
total}(s_{n-1})-t_{\rm total}(s_{n-2})\right]$. The sequence converges
to $s_{\rm equil}$, such { that} $t_{\rm total}(s_{\rm
equil})=0$. We stop when we reach a value for $s$ that makes $|t_{\rm
total}(s)|$ smaller than $10^{-4}t_g${, and we take that as} our
solution for $(\frac{1}{{\Sigma}}\frac{{\rm d}{\Sigma}}{{\rm
d}r})_{\rm equil}$.

With this procedure, we get $(\frac{1}{{\Sigma}} \frac{{\rm
d}{\Sigma}}{{\rm d}r})_{\rm equil}$ for each streamline or,
equivalently, each $r_i(\pi)$. It represents a data point for the
relative radial derivative of the density, shown as a cross on
Fig.~\ref{fig:semi-anal-sol}.

\section*{C: Flux of angular momentum.}

The flux of angular momentum has to be evaluated in a frame in which
angular momentum is conserved. This is not the case for the frame
centered on the primary (which is accelerated), whereas it is the case
in the non-rotating frame centered on the barycenter $G$ of the system
(star plus planet plus disk), which is inertial. One therefore needs
to evaluate the following expression\,:
\begin{equation}
F_H = \int_0^{2\pi} (\Sigma r v_\theta') v_r' r{\rm d}\theta,
\label{eq:F_H_theory}
\end{equation}
where $v_\theta'$ and $v_r'$ are the perturbed azimuthal and radial
velocities in the $G$ centered frame\,:
$v_\theta'=v_\theta-\bar{v_\theta}$ and $v_r'=v_r-\bar{v_r}$, the
barred quantities denoting the averages over the circle of
integration.

We assume that $q\ll 1$. We remark that the perturbed quantities are
proportionnal to $q$, and $F_H$ to $q^2$. Then, to compute
\eqref{eq:F_H_theory} from the velocities output by the code, we need
a sequence of transformations. Neglecting terms that will give
corrections of order $q^3$ in $F_H$, this reduces to two
transformations on $v_r$ and $v_\theta$\,:

\begin{enumerate}
\item[(i)] The velocity of $G$ in the heliocentric frame has to be
substracted. In polar coordinates centred on the star, it is to first
order in $q$\,: $\vec{v}(G)= q
r_p\Omega_p\big(\sin(\theta-\theta_p), \cos(\theta-\theta_p)\big),$
where the subscript $p$ refers to the planet.
\item[(ii)] The radial and azimuthal components of a fluid element
velocity are different in the heliocentric and barycentric frames.
For any vector $X=(X_r,X_\theta)$ in the heliocentric frame, the
radial component in the barycentric frame is written, to first order
in $q$, as\,: $X_r-X_\theta q(r_p/r)\sin(\theta-\theta_p)$.
Similarly, the azimuthal component of $X$ in the barycentric frame
is\,: $X_\theta + X_r q(r_p/r)\sin(\theta-\theta_p)$.  We stress
that the radial component of the velocity of a fluid element is
proportional to $q$, so that the above correction on the azimuthal
component is second order in $q$ and will be neglected.
\end{enumerate}

The application of (i) and (ii) give the following expression for $F_H$\,:
\begin{equation*}
F_H = \bar{\Sigma} \int_0^{2\pi}\big[ r
\left(v_\theta'-qr_p\Omega_p\cos(\theta-\theta_p)\right)\times
\end{equation*}
\begin{equation}
\left(v_r'-q\frac{r_p}{r}(r\Omega_p+\bar{v_\theta})\sin(\theta-\theta_p)\right)\big]
\ r{\rm d}\theta\ ,
\label{eq:F_H_barycentric}
\end{equation}
where $\bar{\Sigma}$ is the mean density on the circle and all the
quantities are the ones output by the code in the heliocentric
frame. This corresponds to the flux of angular momentum through the
circle of radius $r$, due exclusively to the wave launched by the
planet.

The assumption $q\ll 1$ has allowed us to neglect the following
effects\,:
\begin{enumerate}
\item[(a)] The density $\Sigma$ should be evaluated along the circle,
but $\Sigma=\bar{\Sigma}+\Sigma'$, and $\Sigma'\propto
q\Sigma\ll\bar{\Sigma}$, so that $\Sigma$ can be replaced by
$\bar{\Sigma}$ in the integral.
\item[(b)] The circle of radius $r$ centered on $G$ differs from the
circle of radius $r$ centered on the star. As the distance between the
two circles is proportional to q, this only introduces negligible
modifications in the value of every quantity.
\item[(c)] In the previous calculations, $G$ corresponds to the
barycenter of the star-planet system, and not of the whole system
including the disk. The latter is initially axisymmetric, and the
perturbations are proportional to $q$. As the mass of the disk is
also of the order of the mass of the planet, the influence of the disk
on the barycenter position is negligible.
\end{enumerate}

\begin{figure}[t!]
\includegraphics[width=0.7\linewidth,angle=270]{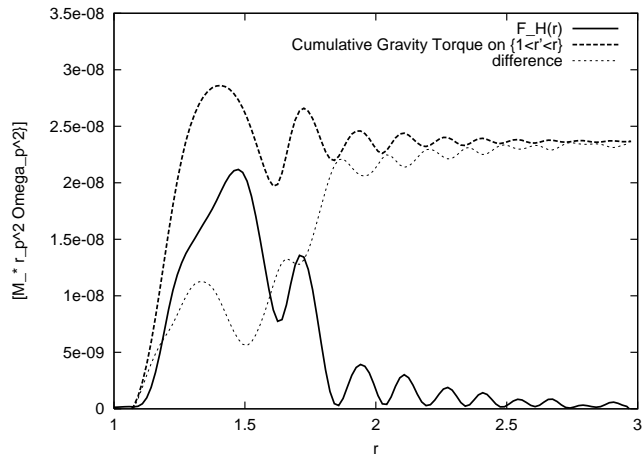}
\caption{\small Angular momentum flux carried by the wave launched by
the planet (bold plain line, corresponding to
Eq.~\eqref{eq:F_H_barycentric}), compared to the total gravity torque
(bold dashed line) as functions of the distance to central star. The
difference is plotted as a thin dot-dashed line.}
\label{fig:flux}
\end{figure}

We computed the flux $F_H$ on our reference simulation ($q=10^{-3}$,
$\nu=10^{-5.5}$, $H/r=0.05$) using Eq.~\eqref{eq:F_H_barycentric}. In
Fig.~\ref{fig:flux}, $F_H(r)$ is plotted as a bold plain line, whereas
the total gravity torque $T_g$ computed on the annulus between the
planet orbit and the circle of radius $r$ is shown as a bold dashed
line. The gravity torque is computed using the direct terms due to the
planet ($GM_p/d^2$, $d$ being the distance between the planet and the
considered point) and to the star ($GM_*/r^2$), as it is evaluated in
an inertial frame. The difference between $F_H$ and $T_g$ is the thin
dot-dashed line\,; it corresponds to the cumulative locally deposited
gravity torque (i. e. the fraction of the gravity torque that is not
evacuated by the pressure supported wave). The wave carries an
increasing flux near the planet (in the zone $\{1.15\lesssim r\lesssim
1.5\}$), and takes away a large fraction of the gravity torque\,; this
corresponds to the raduis where the pressure torque $t_P$ appears to
be of the same order as the gravity torque $t_g$ (see
Fig.~\ref{fig:TpTgTnu_r0}). This angular momentum is then deposited
further from the planet, in particular in the $\{1.5\lesssim r
\lesssim 2\}$ region, where $F_H(r)$ sharply decreases. Beyond $r\sim
2$ the flux vanishes. At the outer boundary of our grid, the flux of
angular momentum taken away by the wave is negligible with respect to
the total gravity torque.

This shows that the outer boundary of the grid is sufficiently far
from the planet so that the angular momentum transfer from the wake to
the disk is correctly described, while the angular momentum leakage at
the outer boundary is negligible. Thus, we conclude that our
simulations are realistic, and our gap profiles correspond to steady
states in the non-migrating planet hypothesis.

\end{document}